\documentclass[review]{elsarticle}

\journal{Journal of Systems and Software}

\pdfoutput=1

\usepackage{amssymb}
\usepackage{times}
\usepackage{url}
\usepackage{graphicx}
\usepackage{amsmath}
\usepackage{booktabs}
\usepackage{diagbox}
\usepackage{algorithm,algpseudocode}
\usepackage{amssymb,amsthm}
\usepackage{subfig}
\usepackage{array}
\usepackage{indentfirst}
\usepackage{url}
\usepackage{enumerate}
\usepackage{multirow}
\usepackage{bbm}
\usepackage{stfloats}
\usepackage{footnote}
\usepackage{color}
\begin{document}

\begin{frontmatter}

\title{Market-Oriented Online Bi-Objective Service Scheduling for Pleasingly Parallel Jobs with Variable Resources in Cloud Environments}

\author{Bingbing Zheng}
\author{Li Pan\corref{cor}}
\ead{panli@sdu.edu.cn}
\author{Shijun Liu\corref{cor}}
\cortext[cor]{Corresponding author}
\ead{lsj@sdu.edu.cn}
\address{School of Software, Shandong University, Jinan, China}

\begin{abstract}
In this paper, we study the market-oriented online bi-objective service scheduling problem for pleasingly parallel jobs with variable resources in cloud environments, from the perspective of SaaS (Software-as-as-Service) providers who provide job-execution services.
The main process of scheduling SaaS services in clouds is: a SaaS provider purchases cloud instances from IaaS providers to schedule end users' jobs and charges users accordingly.
This problem has several particular features, such as the job-oriented end users, the pleasingly parallel jobs with soft deadline constraints, the online settings, and the variable numbers of resources.
For maximizing both the revenue and the user satisfaction rate, we design an online algorithm for SaaS providers to optimally purchase IaaS instances and schedule pleasingly parallel jobs.
The proposed algorithm can achieve competitive objectives in polynomial run-time.
The theoretical analysis and simulations based on real-world Google job traces as well as synthetic datasets validate the effectiveness and efficiency of our algorithm.
\let\thefootnote\relax\footnotetext{© 2021. This manuscript version is made available under the CC-BY-NC-ND 4.0 license http://creativecommons.org/licenses/by-nc-nd/4.0/}
\end{abstract}


\end{frontmatter}


\section{Introduction}
The application scope of cloud computing is getting wider and wider.
In clouds, resources are rent out to users in the form of services with a pay-per-use way, including Infrastructure-as-a-Service (IaaS) and Software-as-a-Service (SaaS), etc. \cite{Armbrust2010A}.
IaaS providers, such as Amazon EC2 \cite{amazon} and Microsoft Azure \cite{microsoft}, offer virtual machines (VMs) with different configurations (including vCPU, memory, storage, etc.) and prices as multiple types of \emph{instances} (e.g., \texttt{c5.large} and \texttt{t3.small}). Most IaaS platforms support pay-as-you-go \emph{on-demand} instances, i.e., one needs to pay per instance per unit time it has used.
SaaS providers sell professional services to users, such as job-execution, web hosting, email access, and so on.
In practice, to reduce the cost of establishing data centers and improve the flexibility of providing services, many SaaS providers tend to elastically purchase on-demand instances from IaaS providers and deploy their services on them.
Considering that the purchase of instances from IaaS providers should be dynamic according to real-time demands, the resources of SaaS providers are \emph{variable}.
By trading SaaS services, on the one hand users only need to care about the service results without considering building data centers and configuring softwares, while on the other hand SaaS providers get revenue from users and pay for IaaS resources. This is featured as a \emph{market-oriented} environment.

\begin{figure}[!t]
\centering
\includegraphics[width=4.2in]{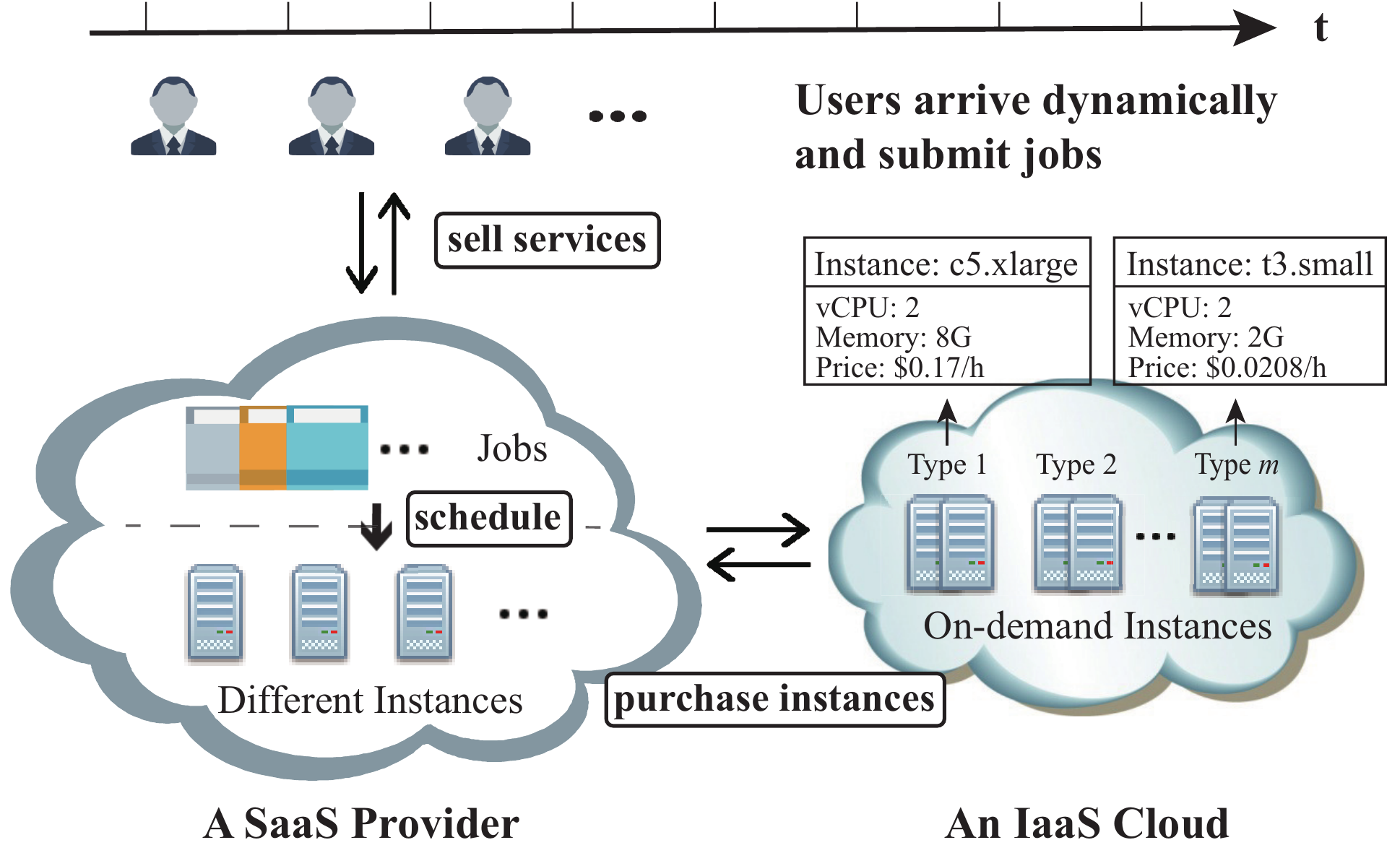}
\caption{The main process for SaaS providers to provide services: a SaaS provider dynamically purchases IaaS instances to execute users' jobs and charges them accordingly.}
\label{model}
\end{figure}

In this paper, we consider SaaS providers who provide job-execution services and focus on \emph{pleasingly parallel jobs} (also called embarrassingly parallel jobs), which can be set flexible degrees of parallelism (DoPs) and divided into any number of tasks to run on different instances in parallel, without any overhead \cite{gunarathne2011cloud}.
In clouds, a large percentage of jobs are pleasingly parallel jobs. Typical examples are image processing such as 3D rendering, scientific computing such as BLAST searches, Monte Carlo simulations, parametric studies, massive searches such as key breaking.
Besides, there are some kinds of data processing and analytics jobs which are pleasingly parallel. For example, map-only jobs such as the Smith Waterman algorithm fit into this type of jobs (while classic map-reduce jobs do not fit because they require synchronization and communications among tasks).
The main process of service scheduling is shown in Fig. \ref{model}. Users arrive over time and submit their pleasingly parallel jobs. A SaaS provider purchases instances from an IaaS provider, schedules the submitted jobs selectively on purchased instances, returns the job-execution results to users, and charges them accordingly.
In this process, SaaS providers should make decisions on the instance purchasing and job scheduling schemes.

In the above scenario, users arrive dynamically and SaaS providers need to make real-time decisions, while the information about the future is generally unknown and hard to be precisely predicted. Meanwhile, current decisions could have good or bad implications for the future.
Therefore, an important issue to care about is that the instance purchasing and job scheduling decisions need to be made in an \emph{online} mode without future information. This is practical in cloud environments but brings more difficulties.
Besides, users in clouds are generally featured as \emph{job-oriented}, which means they only pay attention to the job execution results and ignore the concrete execution process.
Considering the features of pleasingly parallel jobs and job-oriented users, when scheduling jobs, besides the order of executing jobs, proper DoP and instances should also be selected. Different choices result in flexible execution time and completion time, corresponding to the \emph{soft deadline} constraints of jobs.

All along, scheduling problems in various environments are always widely studied.
Traditional environments such as machines \cite{azar2017truthful}, supercomputers \cite{Cao2017Cooling} and clusters \cite{Guo2017iShuffle} do not consider the cost of resources. Their main goal of scheduling is optimizing particular system objectives, including minimizing average makespan and maximizing throughput, etc.
Conversely, market-oriented cloud environments \cite{Buyya2009Cloud} often use commercial models with monetary cost and consider diverse quality of service levels. Thus, market-related objectives (e.g., revenue and cost) become major considerations.
For SaaS providers, \emph{revenue} (i.e., the income earned from users minus the cost paid to IaaS providers) is often the most significant concern.
Meanwhile, to guarantee long-term profits, \emph{user satisfaction rate} (i.e., the rate of the users whose jobs are completed by deadlines) is also an important goal, since rejecting too many users will result in their departure.
Therefore, we deal with \emph{bi-objective} scheduling problems and take both two objectives mentioned above into consideration.

Focusing on the market-oriented online bi-objective service scheduling for pleasingly parallel jobs with variable resources in cloud environments, we design an online algorithm for SaaS providers to help them make better decisions.
In more detail, our contributions are summarized as follows:
(1) We formally define and describe the market-oriented online bi-objective service scheduling problem focusing on pleasingly parallel jobs with various resources, and prove it as being NP-hard (cf. Theorem \ref{thm1}).
(2) We then propose an online algorithm to efficiently decide the instance purchasing and job scheduling, which makes a trade-off between two objectives: maximizing SaaS providers' revenue and maximizing user satisfaction rate.
(3) We utilize strict theoretical analysis to prove that the proposed online algorithm achieves competitive objectives in polynomial run-time. Meanwhile, we construct extensive simulations based on synthetic and real-world Google job data to verify its effectiveness and efficiency.

The rest of the paper is organized as follows. Section 2 introduces the related work. Section 3 gives fundamental notations and formulates the market-oriented online bi-objective service scheduling problem for pleasingly parallel jobs with various resources in clouds as an integer programming. Section 4 designs an online algorithm with strict theoretical analysis. Section 5 constructs extensive evaluations. Section 6 concludes this paper and proposes future research direction.

\section{Related Work} 

Recently, there have been studies on traditional scheduling problems that focus on system objectives.
Azar et al. \cite{azar2017truthful} utilize the correlated rounding technique to design a scheduling algorithm for restricted-related machines, with the aim of minimizing the makespan.
Cao et al. \cite{Cao2017Cooling} present a node location-aware job scheduling algorithm to improve the total throughput for the HPC system. 
Stec et al. \cite{Stec2019Scheduling} solve the stochastic processing time scheduling problem on parallel identical machines to maximize the probability that all the jobs will be completed before a specified deadline.
In \cite{Qi2019Semi}, Qi et al. investigate the problem of scheduling jobs on two identical machines and design a semi-online hierarchical algorithm for minimizing the load of machines.

In market-oriented environments such as clouds, service providers are more concerned about cost or profit.
Li et al. \cite{2016A} focus on the utility value maximization problem for parallel and time-sensitive applications. They design a spatial-temporal interference based scheduling algorithm which is 2-approximate.
Sahni et al. \cite{Sahni2018A} focus on the workflow scheduling problem with deadline constraints in clouds. They design a cost-effective heuristic algorithm from the perspective of an application to  minimize cost.
Alkhanak et al. \cite{Alkhanak2018A} design a completion time driven hyper-heuristic algorithm to help service providers schedule scientific jobs in clouds. Their objective is to optimize the cost of executing jobs.
Unfortunately, all the above algorithms are under offline settings and cannot make decisions without future information.

To address the online mode in cloud environments, some researchers have focused on designing online algorithms.
Zhang et al. \cite{zhang2016online} present a scheduling algorithm from the perspective of cloud brokers that gather resource requests from customers to utilize the volume discount. The algorithm is stack centric, randomized, and online.
Ghose et al. \cite{ghose2017energy} focus on the scientific workflow scheduling problem in cloud systems and design six different scheduling approaches. Their main goal is energy efficiency.
Dambreville et al. \cite{Dambreville2017Load} address the online scheduling of cloud servers to minimize energy consumption.
In \cite{Bao2018Online}, Bao et al. propose an online strategy for scheduling jobs in distributed machine learning clusters. Their objective is to maximize the total utility of executing jobs, which depends on respective completion time of users.
All the above studies take the assumption that resources are fixed as a prerequisite. In fact, adjusting the resources dynamically according to real-time demands is more efficient.

Rare studies design algorithms from the perspective of SaaS providers and take both the scheduling and resource acquiring into consideration at the same time, while simultaneous consideration increases the difficulty significantly.
Wu et al. \cite{Wu2020Toward} design a policy to decide how to purchase on-demand and spot instances for the jobs. The aim is to minimize the cost of purchasing instances.
In another work \cite{zhu2018scheduling}, Zhu et al. consider the workflow scheduling in a hybrid-cloud. The resources are scalable rather than fixed in quantity. They design an iterated heuristic algorithm to reduce the cost of renting VMs.
But these two studies only consider offline settings. 
Cao et al. \cite{Cao2017Online} propose an online algorithm which schedules service requests in hybrid clouds to minimize the cost of renting resources from public clouds. 
However, all the above researches have single objectives.
Instead, we consider more complex objectives including revenue and user satisfaction rate, which conforms better to the practice with the long-term development considerations.

There are also many studies that consider multiple objectives when scheduling jobs.
Huang et al. \cite{huang2018multi} address the task assignment problem in clouds and develop a new algorithm for multi-objective scheduling on the basis of particle swarm optimization, with aims of minimizing power consumption and makespan.
Kaur et al. \cite{kaur2018a} propose a novel multi-objective job scheduling strategy in cloud and grid environments. In their paper, they make a trade-off among flowtime, makespan, and resource cost.
In \cite{peng2019joint}, Peng et al. solve the task scheduling problem in mobile cloud environments. They design a scheduling algorithm on the basis of dynamic voltage scaling technique and whale optimization strategy to jointly optimize completion time and energy consumption. 
These three algorithms are offline algorithms which are unsuitable for online settings.
Hu et al. \cite{hu2018time} propose an online task scheduling strategy to optimize both completion time and costs in geographically distributed data centers.
Gasior et al. \cite{gasior2017a} present a framework of studying multi-objective online scheduling problem in IaaS cloud computing systems. Their objectives include minimizing the SLA violation counts and maximizing the total provider's income.  
Nevertheless, these two online multi-objective algorithms do not consider the dynamic purchasing of resources.

Particularly, focusing on pleasingly parallel jobs, there are many studies that have been proposed.
Gunarathne et al. \cite{2011Cloud} present two pleasingly parallel biomedical applications and use them to analyze the performance of different cloud models.
\cite{2016A} mentioned above focuses on parallel and time-sensitive applications, which are in line with pleasingly parallel jobs.
Zhang et al. \cite{Zhang2014Evolutionary} focus on scheduling the multitasking workloads for big-data analytics. They introduce the ordinal optimization using rough models and the fast simulation to get the suboptimal solutions at a very fast speed.
Stavrinides et al. \cite{Stavrinides2017The} concentrate on the big data analytics SaaS and study data-aware scheduling policies.
Chen et al. \cite{Chen2019Scheduling} address the scheduling problem of big data analytics jobs. Their goal is to minimize the total job completion time while achieving max-min fairness.
These algorithms are all offline algorithms.
Shi et al. \cite{2017Energy} solve the problem of scheduling pleasingly parallel jobs and design an online algorithm to minimize job completion time.
Zhou et al. \cite{Zhou2017An} focus on parallel computing jobs with soft deadline constraints and utilize the technique of primal-dual to design an online algorithm.
The two algorithms mentioned above are online algorithms which do not consider variable resources.
In \cite{Wu2020Toward}, Wu et al. design a policy to allocate self-owned, on-demand, and spot instances to the arriving pleasingly parallel jobs, using CAP3 in bioinformatics as an example. Their goal is cost optimization. The resources in this work are variable but the policy is offline.
In summary, none of the above scheduling algorithms considers the pleasingly parallel jobs, the online settings, the variable resources, and the complex objectives at the same time.

After summarizing the existing work, the main contribution of this paper lies in that we design an online service scheduling algorithm for pleasingly parallel jobs with various resources in market-oriented cloud environments. In this algorithm, we take both the SaaS providers' revenue and user satisfaction rate into consideration and help SaaS providers efficiently solve their scheduling problems.

\section{System Model}
In this section, we first introduce the fundamental notations used in this paper and then use these notations to formulate the market-oriented online bi-objective service scheduling problem for pleasingly parallel jobs.

\subsection{Fundamental Notations}

The SaaS providers periodically sum up the income and cost, and a period is called an accounting period.
The time axis of an accounting period consists of $T$ discrete time slots, represented as $\mathcal{T}\! =\!\{1, 2, \ldots, T\}$. User arrivals, instance purchasing, and job scheduling are based on time slots.
In this paper, we only consider on-demand IaaS instances whose usage duration is $\tau$ slots.
We assume that SaaS providers purchase $m$ types of on-demand instances from a public IaaS cloud, written as $\mathcal{M}\!=\!\{1, 2,\ldots, m\}$.
These different types of heterogeneous instances have different physical configurations (including CPU and memory etc.) and correspond to different purchase prices.
$\mathit{price_{j}}$ represents the price of type-$j$ instances, which is the money a SaaS provider needs to pay if it purchases an instance $j$ unit. 
An instance $j$ unit corresponding to a type-$j$ instance running for one time slot.
$r_{j}(t)$ denotes the number of type-$j$ instances newly purchased at time slot $t$. Considering the huge capacity of public IaaS clouds, in theory $r_{j}(t)$ can be infinite.
We use $c_{j}(t)$ to denote the number of type-$j$ instances that are available at time slot $t$.

We use $n$ to denote the total number of users who arrive in the whole accounting period, written as $\mathcal{U}\! =\!\{1, 2,\ldots, n\}$. User $i$ arrives at time slot $a_{i}$ and submits a pleasingly parallel job to the SaaS provider. For simplicity and without ambiguity, we also use $i$ to denote the job of user $i$.
Only when a job arrives at the system, its information can be known by SaaS providers.
Each job $i$ can be described by a tuple ($a_{i}$, $d_{i}$, $\{D_{i,j}\}_{j\in\mathcal{M}}$, $k_{i}$, $v_{i}$).
$a_{i}$ is the arrival time and $d_{i}$ is the deadline. Job $i$ can be completed at any time slot between $a_{i}$ and $d_{i}$. 
We use $D_{i,j}$ to represent the demand of job $i$ for type-$j$ instances. It means to complete job $i$, total $D_{i,j}$ instance $j$ units are required. The demands of job $i$ for different types of instances are related to instances' performances. A job can be executed on multiple types and numbers of instances, as long as the total demand is satisfied.
Estimating a job's demand which can also be understood as the runtime is a classic problem which attracts extensive studies (e.g., \cite{Chen2019PISCES, Liang2015A}), falling beyond the scope of our study.
Threshold $k_{i}$ is used to limit job $i$'s degrees of parallelism (DoPs), which means job $i$ can be divided into at most $k_{i}$ independent tasks to run on different instances in parallel, without any overhead.
$v_{i}$ represents the value function of user $i$, which is related to the execution time of job $i$. This value function also denotes user $i$'s willingness to pay. Generally, $v_{i}$ will decrease as the execution time increases.
In this paper, we use $i$'s value function to indicate the valuation that SaaS providers can obtain from executing job $i$. As a result, SaaS providers charge a corresponding amount of money as payment.
The value functions can be arbitrary shapes as long as they are nonincreasing functions.
For instance, the value function can be a linear decreasing function or have two stages (e.g., a constant until the deadline and a linear decreasing function after the deadline).
Besides, since we do not consider the penalties of SaaS providers, the value function should be no less than 0.

We use $e$ to represent the execution time which is the number of time slots that SaaS providers spend to execute a job. Considering the feature of jobs and users, with different scheduling schemes, the execution time is different. For job $i$, considering its arrival time $a_{i}$ and deadline $d_{i}$, we have its execution time $e\in \mathcal{E}_{i}$, where $\mathcal{E}_{i}\!=\!\{1, 2,\ldots, d_{i}-a_{i}\}$. Then the corresponding completion time is $a_{i}+e$, and the corresponding valuation is $v_{i}(e)$.
To express discrete time axis and make it convenient to correspond, we use $v_{i}^{e}$ to represent the valuation of user $i$ if its execution time is $e$.
We use binary variable $x_{i}^{e}$ for indicating whether job $i$'s execution time is $e$. 
Job $i$ may have multiple possible values of execution time $e\in\mathcal{E}_{i}$. Thus, the execution time $e$ which makes $x_{i}^{e}=1$ is its actual execution time.
If a job's execution time is $e$, we also say that job $i$ with execution time $e$ is accepted.
Then the values of $x_{i}^{e}$ is:
\[
x_{i}^{e}=\left\{
\begin{array}{lcl}
1       &      & {\mathit{\text{if job $i$'s execution time is $e$}}}\\
0       &      & {\mathit{\text{otherwise}}}
\end{array}
\right.\]
The payment of user $i$ that the SaaS provider should charge is
$p_{i}=\sum_{e\in\mathcal{E}_{i}} v_{i}^{e} x_{i}^{e}$, which equals the actual valuation of job $i$.
In this paper, we assume that SaaS providers can reject service requests without any penalties.
For reference, Table \ref{symbol} summarizes the fundamental notations that are used in this paper.

To provide services, SaaS providers need to make the following decisions:\\
(1) The number of type-$j$ instances that should be newly purchased from an IaaS cloud at each time slot $t$, $r_{j}(t)$.\\
(2) Whether to accept user $i$'s job with execution time $e$, $x_{i}^{e}$.\\
(3) The number of type-$j$ instances allocated to user $i$ at time slot $t$ with execution time $e$, $y_{i,j}^{e}(t)$.

Considering the features of pleasingly parallel jobs, job $i$ can be scheduled with different DoPs on multiple types and numbers of instances, which results in diverse execution time and ulteriorly leads to different valuations to user $i$.
For better understanding, we give a brief example.
Job $i$ arrives at time slot $a_{i}=1$ with deadline $d_{i}=5$. Then the set of possible execution time is $\mathcal{E}_{i}={1, 2, 3, 4}$.
It has $D_{i,1}=4$, $D_{i,2}=2$ and $k_{i}=5$.
There are several possible scheduling schemes.\\
(1) Job $i$ can be executed on 4 type-$1$ instances for 1 time slot (i.e., DoP is 4). Then $e=1$ and completion time is 2. $y_{i,1}^{1}(t)=4, t=2$. The corresponding valuation is $v_{i}^{1}=100$.\\
(2) Job $i$ can be executed on 1 type-$1$ instance for 4 time slots (i.e., DoP is 4). Then $e=4$ and completion time is 5. $y_{i,1}^{4}(t)=1, \forall t\in[2, 5]$. The corresponding valuation is $v_{i}^{4}=30$.\\
(3) Job $i$ can be executed on 1 type-$2$ instance for 1 time slot and 1 type-$1$ instances for 2 time slots (i.e., DoP is 3). Then $e=2$ and completion time is 3. $y_{i,2}^{2}(t)=1, t=2$ and $y_{i,1}^{2}(t)=1, \forall t\in[2, 3]$. The corresponding valuation is $v_{i}^{2}=80$.\\
\quad Or, $e=3$ and completion time is 4. $y_{i,2}^{3}(t)=1, t=2$ and $y_{i,1}^{3}(t)=1, \forall t\in[3, 4]$. The corresponding valuation is $v_{i}^{3}=50$.

\begin{table}
\renewcommand{\arraystretch}{1.5}
\caption{Summary of Fundamental Notations}
\label{symbol}
\newcommand{\tabincell}[2]{\begin{tabular}{@{}#1@{}}#2\end{tabular}}
\centering
\begin{tabular}{|p{1cm}<{\centering}|c|p{1cm}<{\centering}|c|}
\hline
$T$ & number of time slots & $\mathcal{T}$ & set of time slots\\
\hline
$n$ & number of users & $\mathcal{U}$ & set of users\\
\hline
$a_{i}$ & arrival time of user $i$ & $d_{i}$ & deadline of user $i$\\
\hline
$p_{i}$ & payment of user $i$ & $\mathcal{M}$ & set of instance types\\
\hline
$m$ & number of instance types & $\tau$ & usage duration of instances \\
\end{tabular}

\begin{tabular}{|p{1.2cm}<{\centering}|p{9cm}<{\centering}|}
\hline
$r_{j}(t)$ & number of newly purchased type-$j$ instances at time slot $t$ \\
\hline
$price_{j}$ & price of purchasing one unit type-$j$ instance for one time slot \\
\hline
$\mathcal{E}_{i}$ & set of possible execution time of job $i$ \\
\hline
$D_{i,j}$ &  demand of job $i$ for type-$j$ instances  \\
\hline
$k_{i}$ & threshold of job $i$'s degrees of parallelism (DoPs)\\
\hline
$v_{i}^{e}$ & valuation of job $i$ with execution time $e$\\
\hline
$x_{i}^{e}$ &  whether job $i$'s execution time is $e$ (if yes $x_{i}^{e}=1$, otherwise $x_{i}^{e}=0$)\\
\hline
$y_{i,j}^{e}(t)$ & \tabincell{c}{number of type-$j$ instances allocated to job $i$ \\ with execution time $e$ at time slot $t$}\\
\hline
$c_{j}(t)$ & number of available type-$j$ instances at time slot $t$\\
\hline
$\phi_{i}^{e}$ & cost of user $i$ with execution time $e$\\
\hline
$\hat{\phi_{i}^{e}}$ & estimated offline cost of user $i$ with execution time $e$ \\
\hline
\end{tabular}
\end{table}

\subsection{Problem Formulation}
Assuming that all the required information is known in advance, we use the fundamental notations mentioned above to formulate the market-oriented online service scheduling problem with various resources in cloud environments.

To build long-term attraction, when solving the scheduling problem, SaaS providers need to consider two objectives:\\
(1) Maximizing SaaS providers' revenue ($\mathit{REV}$): The revenue of a SaaS provider is the total payment charged to users (i.e, the total valuation of job completions) minus the total cost of purchasing instances. SaaS providers pursue the greatest revenue as much as possible.
\begin{equation}\label{REV}
\text{max} \quad \mathit{REV}=\sum_{i\in\mathcal{U}}\sum_{e\in \mathcal{E}_{i}} v_{i}^{e} x_{i}^{e}-\sum_{j\in\mathcal{M}}\sum_{t\in\mathcal{T}}\mathit{price_{j}} r_{j}(t)
\end{equation}\\
(2) Maximizing user satisfaction rate ($\mathit{SAT}$): If a user's job is completed before its deadline, then this user's service request is satisfied, which also means this user is satisfied.
We define the user satisfaction rate of the system as the rate of satisfied users.
Thus, to maximize the user satisfaction rate, we need to maximize the rate of the users whose jobs are completed by deadlines. According to the definition, $\mathit{SAT}\in[0, 1]$.
Guaranteeing a high user satisfaction rate can prevent users from leaving, which is beneficial to long-term profit.
\begin{equation}\label{SAT}
\text{max} \quad \mathit{SAT}=\frac{\sum_{i\in\mathcal{U}}\sum_{e\in\mathcal{E}_{i}}x_{i}^{e}}{n}
\end{equation}

Optimizing the above two objectives simultaneously belongs to the multi-objective optimization problem. One classic method to address this problem is to transform several objectives into one aggregated objective. To do this, one can use the weighted sum method \cite{1963Optimality, 2010The}, which calculates the linear combination of several different objectives.
In this paper, to take both the revenue $\mathit{REV}$ and user satisfaction rate $\mathit{SAT}$ into consideration, we use the idea of the weighted sum method and linearly combine these two objectives by a factor $\theta$ to get an aggregated objective $\mathit{OBJ}$. The factor $\theta$ indicates the relative importance of objectives $\mathit{REV}$ and $\mathit{SAT}$. It can be set by SaaS providers.
Considering that the ranges of $\mathit{REV}$ and $\mathit{SAT}$ are very different, we normalize these two objectives by min-max normalization \cite{2006Data}. 
In this way, these two objectives can be scaled to fall within a specified range (i.e., $[0, 1]$) and thus generate comparable influence to the final aggregated objective $\mathit{OBJ}$.
\begin{equation}\label{OBJ}
\text{max} \quad \mathit{OBJ}=\theta\cdot\frac{\mathit{REV}}{\mathit{REV^{*}}}+(1-\theta)\cdot \mathit{SAT}
\end{equation}
Here $\mathit{REV^{*}}$ is the optimal revenue the system can obtain in an accounting period.

Then we can formulate the market-oriented online service scheduling problem for pleasingly parallel jobs with various resources in cloud environments to an integer programming (IP):
\allowdisplaybreaks
\begin{alignat}{2}\label{IP}
\text{max} \quad  &\theta\cdot\frac{\mathit{REV}}{\mathit{REV^{*}}}+(1-\theta)\cdot \mathit{SAT} \\
\text{s.t.}\quad
&\sum_{j=1}^{m}\sum_{t=a_{i}+1}^{a_{i}+e} y_{i,j}^{e}(t) \frac{1}{D_{i,j}}\geq x_{i}^{e}  &\ & \forall\ i\in\mathcal{U}, e\in \mathcal{E}_{i} \label{c1}\\
&\sum_{e=1}^{d_{i}-a_{i}} x_{i}^{e} \leq 1  &\quad & \forall\ i\in\mathcal{U}\label{c2}\\
&\sum_{j=1}^{m}\sum_{t=a_{i}+1}^{a_{i}+e} y_{i,j}^{e}(t) \leq k_{i}  &\quad & \forall\ i\in\mathcal{U}, e\in \mathcal{E}_{i}\label{c3}\\
&c_{j}(t)=\sum_{t'=t-\tau+1}^{t}r_{j}(t')  &\quad & \forall\ t\in\mathcal{T}, j\in\mathcal{M}\label{c4}\\
&\sum_{i=1}^{n}\sum_{e=1}^{d_{i}-a_{i}} y_{i,j}^{e}(t)\leq c_{j}(t)  &\quad & \forall\, j\in\mathcal{M}, t\in\mathcal{T} \label{c5}\\
&x_{i}^{e}\in \{0,1\} &\quad & \forall\ i\in\mathcal{U}, e\in \mathcal{E}_{i}\label{c6} \\
&y_{i,j}^{e}(t) \in \mathbb{N} &\quad & \forall\ i\in\mathcal{U}, e\in \mathcal{E}_{i}, j\in\mathcal{M}, t\in\mathcal{T} \label{c7} \\
&r_{j}(t) \in \mathbb{N} &\quad & \forall\,j\in\mathcal{M},t\in\mathcal{T} \label{c8}
\end{alignat}Constraint (\ref{c1}) implies that the number of instances distributed to a job should be enough to complete the job. In other words, partial completion is meaningless.
Constraint (\ref{c2}) means one job can be completed at most once.
In practice, practical thresholds should exist to limit DoPs, resulting in constraint (\ref{c3}).
Constraints (\ref{c4}) and (\ref{c5}) are resource capacity constraints.
Constraints (\ref{c7}) and (\ref{c8}) are the integer constraints, requiring the number of instances allocated to jobs and purchased from IaaS clouds at each time slot to be an integer.

\newtheorem{thm}{Theorem}
\begin{thm}\label{thm1}
The market-oriented bi-objective online service scheduling problem for pleasingly parallel jobs with various resources in cloud environments (formulated to IP (\ref{IP})) is NP-hard.
\end{thm}

\begin{proof}
To prove Theorem \ref{thm1}, we reduce our scheduling problem to the 0-1 knapsack problem. This problem is a typical optimization problem which has been proven to be NP-hard \cite{Kellerer2004Knapsack}.
In the reduction, the jobs are considered as items, and the resources are considered as knapsacks.
Focusing on the bi-objective service scheduling problem (\ref{IP}), we do the modifications as follows.
(1) The objective is SaaS providers' revenue maximization, which means equation (\ref{IP}) is modified to $\text{max} \ \mathit{REV}=\sum_{i\in\mathcal{U}}\sum_{e\in \mathcal{E}_{i}} v_{i}^{e} x_{i}^{e}-\sum_{j\in\mathcal{M}}\sum_{t\in\mathcal{T}}\mathit{price_{j}} r_{j}(t)$. 
(2) Only one time slot is considered ($|\mathcal{T}|=1$), which means symbol $t$ is omitted. Correspondingly, all jobs come and complete at the same time (
$|\mathcal{E}_{i}|=1$), which means symbol $e$ is omitted.
(3) Only one type of instances is used ($|\mathcal{M}|=1$), which means symbol $j$ is omitted. In combining with (2), there is only one scheduling scheme ($y_{i}=D_{i}$) which will not violate the DoP threshold. Thus constraints (\ref{c1}) and (\ref{c3}) are removed.
(4) Dynamic instance purchasing is not considered and the resource capacity of the SaaS provider is fixed,  
which means symbol $r$ is omitted and thus constraints (\ref{c4}) and (\ref{c8}) are removed.
Then our scheduling problem can be reduced as:
\allowdisplaybreaks
\begin{alignat}{2}
\max\quad &\sum_{i\in\mathcal{U}}v_{i} x_{i} \nonumber\\
\mbox{s.t.}\quad
&\sum_{i\in\mathcal{U}}D_{i}\leq c &\quad & \nonumber \\
&x_{i}\in \{0,1\} &\quad & \forall\ i\in\mathcal{U} \nonumber
\end{alignat}From this reduction, the market-oriented online service scheduling problem in this paper can be regarded as a complicated expansion of 0-1 knapsack problem. Consequently, our problem is also NP-hard.

\end{proof}

Offline optimal solutions can be obtained by solving IP (\ref{IP}). 
However, according to Theorem \ref{thm1}, finding exact solutions is NP-hard.
Moreover, online settings should also be considered. When making decisions, SaaS providers should only use the information already known.
Considering these challenges, we propose an online algorithm to make approximate decisions.

\section{An Online Service Scheduling Algorithm}
In this section, we propose an online algorithm and prove its effectiveness through theoretical analysis.
The main steps of the online algorithm are shown in Algorithm \ref{online}.

\begin{algorithm}
    \caption{An Online Service Scheduling Algorithm}\label{online}
    \begin{algorithmic}[1]
        \State When user $i$ arrives at the system
        \For{all $e\in \mathcal{E}_{i}$}
            \State Calculate the instance purchasing scheme $r_{j}(t)$ and job scheduling scheme $y_{i,j}^{e}(t)$ by the job scheduling strategy.
            \State Calculate the cost $\hat{\phi_{i}^{e}}$ of completing job $i$ with execution time $e$ by the cost calculating strategy.
        \EndFor
        \State Let $e^{*}=\arg\max_{e}\{\theta\cdot \frac{v_{i}^{e}-\hat{\phi_{i}^{e}}}{\mathit{REV^{*}}}+(1-\theta)\cdot\frac{1}{n}\}$ be job $i$'s optimal execution time.
        \If {gain increment $\theta\cdot \frac{v_{i}^{e^{*}}-\hat{\phi_{i}^{e^{*}}}}{\mathit{REV^{*}}}+(1-\theta)\cdot\frac{1}{n}\geq 0$}
            \State Accept job $i$ with execution time $e^{*}$.
            \State $x_{i}^{e^{*}} = 1$.
            \State $p_{i}=v_{i}^{e^{*}}$.
            \State Use the job scheduling strategy to re-compute $r_{j}(t)$ and $y_{i,j}^{e}(t)$, according to all the actually accepted jobs.
        \Else
            \State Reject user $i$.
        \EndIf
    \end{algorithmic}
\end{algorithm}

\subsection{The Job Scheduling Strategy}
When user $i$ arrives at the system and submits its service request, the SaaS provider needs to make decisions immediately. In this process, we temporarily assume that all the arrived jobs until now have been accepted. The respective actual execution time is $e^{*}$.
First, we propose a job scheduling strategy which is straightforward and efficient to decide the instance purchasing and job scheduling schemes for job $i$ (line 3 in Algorithm \ref{online}).

In the following description, we take type-$j$ instances as an example.
For job $i$'s each possible execution time $e\in\mathcal{E}_{i}$, we averagely allocate the total demand $D_{i,j}$ to the whole execution interval $[a_{i}+1, a_{i}+e]$, i.e, $y_{i,j}^{e}(t)=\frac{D_{i,j}}{e}, \forall t\in [a_{i}+1, a_{i}+e]$. That is, the number of type-$j$ instances needed by job $i$ is the same in every time slot of interval $[a_{i}+1, a_{i}+e]$.
For each time slot $t$ in $[a_{i}+1, a_{i}+e]$, if the instances which are already purchased in previous time slots are enough to complete corresponding demands, then we allocate these instances to job $i$. Otherwise, we purchase new instances until the instances are sufficient and set $r_{j}(t)$ to the corresponding value.
When new instances are needed, we choose to purchase the instance type which has the maximum performance/price ratio.
In this process, we need to note the DoP threshold. If an instance type cannot guarantee the threshold, then this type of instances will not be considered for executing this job.
As a straightforward method, this job scheduling strategy can efficiently and fast produce feasible schemes.

Now we give a simple example to help understand the job scheduling strategy.
Job $i$'s arrival time $a_{i}=1$, deadline $d_{i}=4$ and demand $D_{i,j}=6$. Its possible execution time is $e={1, 2, 3}$.
For execution time $e=1$, $y_{i,j}^{1}(t)=6, t=2$.
For execution time $e=2$, $y_{i,j}^{2}(t)=3, \forall t\in [2, 3]$.
For execution time $e=3$, $y_{i,j}^{3}(t)=2, \forall t\in [2, 3, 4]$.

\subsection{The Cost Calculating Strategy}
Then, for the obtained job scheduling and instance purchasing schemes, we propose the following cost calculating strategy to calculate the cost of executing job $i$ with execution time $e$ (line 4 in Algorithm \ref{online}). We use $i^{e}$ to represent that job $i$ is executed with execution time $e$.

Since an instance may be shared by multiple jobs, the cost of purchasing this instance should also be distributed among these jobs.
Then we use the idea of proportional sharing \cite{Si2013Proportional} to distribute the total cost among all the accepted users.
Proportional sharing is a simple but effective cost distribution approach which distributes the cost of a jointly used common resource among its users. This method is also called average cost pricing in \cite{Moulin2002Axiomatic, Wang2002Ordinal, Lin2015A}.
The main idea of this method is: dividing the total cost in proportion to individual revenue/demand.
That is, a user's sharing cost is the total cost times the proportion of its revenue/demand to the total revenue/demand.

Theoretically, users' value functions can represent their revenues. However, value functions are personally defined which cannot adequately reflect the cost to complete jobs. Some users may have large valuations for small demands or vice versa. Hence, to distribute the cost fairly, we use the demands to calculate the sharing costs.
However, the other problem is that the cost produced by executing a job is related to not only demand but also execution time.
Thus, when considering the cost distribution of job $i^{e}$, we need to synthesize the demand and execution time. 
To combine demand and execution time, we use the weighted sum method and min-max normalization again mentioned above. The aggregated result is defined as the weight of $i^{e}$ which is used to calculate the sharing cost in equation (\ref{sharing cost}). 
The weight of job $i$ with execution time $e$ is:
\[
w_{i}^{e}=\alpha\frac{D_{i,j}-D_{\mathit{min}}}{D_{\mathit{max}}-D_{\mathit{min}}}+(1-\alpha)\frac{(E-e)-1}{E-1}
\]
where $D_{\mathit{min}}=\min_{i}{D_{i,j}}$, $D_{\mathit{max}}=\max_{i}{D_{i,j}}$, while $E$ is the maximum number of possible execution time. Coefficient $\alpha$ can indicate the importance of demands and execution time. It is chosen from $[0, 1]$ and can be set by SaaS providers based on their requirements.

Then, $i^{e}$'s sharing cost is the total cost times the proportion of $i^{e}$'s weight to the total weight of all accepted users.
The weight $w_{i}^{e}$ here can be seen as the contribution of executing $i^{e}$ to the total cost. Then the sharing cost of $i^{e}$ is:
\begin{equation}\label{sharing cost}
\phi_{i}^{e}= f_{i^{e}}(\mathcal{U}) \frac{w_{i}^{e}}{\sum_{j\neq i}w_{j}^{e^{*}}+w_{i}^{e}}
\end{equation}
$f$ represents the cost function and $f(\mathcal{U})$ represents the total cost of executing all the users in $\mathcal{U}$ with their optimal execution time $e^{*}$.
If we appoint that job $i$'s execution time is $e$, the total cost is represented as $f_{i^{e}}(\mathcal{U})$.

Nevertheless, only if the total cost is known, the cost of each job can be calculated accurately.
In the practical online markets, it is impossible to calculate each user's accurate sharing cost real-timely.
Thus we consider estimating the cost for a user in the whole accounting period.
The main process is as follows.
When the instance purchasing and job scheduling schemes are determined, at first we calculate current cost $\phi_{i}^{e}$ for job $i$ with execution time $e$. The total cost used in this calculation is the current total cost, supposing all the arrived jobs are accepted.
Then, based on $\phi_{i}^{e}$, we apportion job $i$'s cost to the jobs who arrive later than $i$, and apportion the cost produced by the jobs who arrive later than $i$ to job $i$.
That is, the estimated cost $\hat{\phi_{i}^{e}}$ of job $i$ with execution time $e$ in the whole accounting period is:
\[
\hat{\phi_{i}^{e}}=\frac{\phi_{i}^{e}f(\mathcal{U})a_{i}}{f_{i^{e}}(\mathcal{U}_{i})T}
\]
Here, $\mathcal{U}_{i}$ is the set of users who arrive before $i$. $f(\mathcal{U})$ can be obtained using historical data.
In this estimation, we suppose that the probability of job arrivals and demands in every time slot is the same.
By using such an estimation approach, we can conclude:
$E[\sum_{i \in \mathcal{U}} \hat{\phi_{i}^{e^{*}}}]=f(\mathcal{U})$.

\subsection{The Acceptance Strategy}
Next, we use the following acceptance strategy to decide the acceptance or rejection of user $i$ (lines 6-14 in Algorithm \ref{online}).
We choose the execution time which maximizes the \emph{gain increment} (equation \ref{increment}) as the optimal execution time $e^{*}$ of job $i$.
Only if the gain increment of $e^{*}$ is high enough, job $i$ will be accepted and the execution time is $e^{*}$.
If $e^{*}$ is accepted, the payment user $i$ should pay is its valuation corresponding to execution time $e^{*}$.
Finally, we run the job scheduling strategy once again for accepted job $i$ with its execution time $e^{*}$ to calculate actual schemes, according to all the jobs which are actually executed.

The gain increment to the aggregated objective $\mathit{OBJ}$ of job $i$ with execution time $e$ is:
\begin{equation}\label{increment}
\theta \cdot\frac{v_{i}^{e}-\hat{\phi_{i}^{e}}}{\mathit{REV^{*}}}+(1-\theta)\cdot\frac{1}{n}
\end{equation}
Both $\mathit{REV^{*}}$ and $n$ can be obtained according to the historical data of the previous accounting period.

This gain increment considers both SaaS providers' revenue and user satisfaction rate at the same time.
According to $v_{i}^{e}-\hat{\phi_{i}^{e}}$, only the users whose valuations are large enough can get served.
By adding $\frac{1}{n}$, the online algorithm accepts more users whose valuations are small, which will decrease SaaS providers' revenue but increase user satisfaction rate.
$\theta\in[0, 1]$ reflects the importance of providers' revenue and user satisfaction rate. It can be set by SaaS providers according to their requirements.
If $\theta=1$, the goal of the algorithm is revenue maximization. With less $\theta$, the algorithm considers more about user satisfaction rate.

\subsection{Theoretical Analysis}

The online algorithm should achieve good performance and near-optimal objectives. \emph{Competitive ratio} is used to measure the performance of an online algorithm.
If whatever the input is, the ratio of the offline optimal objective to the objective achieved by the online algorithm is smaller than or equal to a specific value $\epsilon$, then the online algorithm can be seen as achieving competitive ratio $\epsilon$.
This section proves that our online algorithm achieves competitive revenue and aggregated objective.

\begin{thm} \label{thm3}
The proposed online algorithm can achieve an expected competitive ratio of $\frac{\delta}{\rho-2}$ in the SaaS providers' revenue $\mathit{REV}$. $\delta$ and $\rho$ are related to users' valuations.
\end{thm}

\begin{proof}
We use subscripts $_{opt}$ and $_{online}$ to represent the optimal solution and the solution obtained by our online algorithm respectively. For instance, $\mathit{REV_{opt}}$ and $\mathit{REV_{online}}$ represent the $\mathit{REV}$ obtained by the optimal algorithm and our online algorithm respectively.

First, we suppose a fictional situation to give an upper bound of both $\mathit{REV}$ and $\mathit{SAT}$: all the jobs are completed at the first time slot after their arrivals with instance purchase cost 0. Apparently, in our problem there is no situation that can produce larger revenue and user satisfaction rate than this fictional situation, including the optimal solution.
Thus we have:
\begin{align}
&\mathit{REV_{opt}}\leq\mathit{REV^{*}}\leq\sum_{i\in \mathcal{U}}v_{i}^{\mathit{max}}\label{rev_opt}\\
&\mathit{SAT_{opt}}\leq 1 \label{sat_opt}
\end{align}
$v_{i}^{\mathit{max}}$ represents user $i$'s maximum valuation. Considering that value functions are nonincreasing, a user's maximum valuation is generally its valuation for the first time slot.

Considering the fictional situation, the upper bound of the $\mathit{REV_{opt}}$ is (equation $\ref{rev_opt}$):
\[\mathit{REV_{opt}} \leq \sum_{i\in \mathcal{U}}v_{i}^{\mathit{max}}\]

Then, we consider the situation using our online algorithm.
Let $\mathcal{W}$ and $\mathcal{L}$ denote the set of accepted and rejected users respectively.
The total numbers of users of these two sets are $n_{1}$ and $n_{2}$ respectively.
$f(\mathcal{W})$ is the actual total cost according to our algorithm. Apparently, this actual cost will be no larger than the total cost of serving all users in $\mathcal{U}$, that is, $ f(\mathcal{W}) \leq f(\mathcal{U})$.
The expectative revenue obtained by our online algorithm is the total valuation of completed jobs minus the actual cost, i.e.:
\begin{align}
E[\mathit{REV_{online}}]&=E[\sum_{i\in \mathcal{W}}v_{i}^{e^{*}} - f(\mathcal{W})]\label{r1}\\
&\geq E[\sum_{i\in \mathcal{U}}v_{i}^{e^{*}} - \sum_{i\in \mathcal{L}}v_{i}^{e^{*}} - \sum_{i\in \mathcal{U}}\hat{\phi_{i}^{e^{*}}}]\label{r2}\\
&\geq E[\sum_{i\in \mathcal{U}}v_{i}^{e^{*}} - \sum_{i\in \mathcal{L}} (\hat{\phi_{i}^{e^{*}}}-\mathit{REV^{*}}\frac{1-\theta}{\theta}\frac{1}{n})-\sum_{i\in\mathcal{U}}\hat{\phi_{i}^{e^{*}}} ]\label{r3}\\
&\geq E[\sum_{i\in \mathcal{U}}v_{i}^{e^{*}}- 2\sum_{i\in \mathcal{U}}\hat{\phi_{i}^{e^{*}}}+\mathit{REV^{*}}\frac{1-\theta}{\theta}\frac{n_{2}}{n}]\label{r4}\\
&\geq E[\sum_{i\in \mathcal{U}}v_{i}^{\mathit{min}}- 2\sum_{i\in \mathcal{U}}\hat{\phi_{i}^{e^{*}}}] \label{r5}
\end{align}
The set of all users $\mathcal{U}$ can be divided into the set of accepted users $\mathcal{W}$ and the set of rejected users $\mathcal{L}$. Besides, $ f(\mathcal{W}) \leq f(\mathcal{U})$ and $E[\sum_{i \in \mathcal{U}} \hat{\phi_{i}^{e^{*}}}]=f(\mathcal{U})$.
Thus we can get equation (\ref{r2}) from equation (\ref{r1}).
According to the acceptance strategy in Section 4.3, the users who are rejected have: $\theta\cdot \frac{v_{i}^{e^{*}}-\hat{\phi_{i}^{e^{*}}}}{\mathit{REV^{*}}}+(1-\theta)\cdot\frac{1}{n}< 0$.
Thus we can get equation (\ref{r3}) from equation (\ref{r2}).
Since the cost of rejected users must no larger than the total cost of serving all users, we can get equation (\ref{r4}) from equation (\ref{r3}).
Let $v_{i}^{\mathit{min}}$ represent user $i$'s minimum valuation. Then equation (\ref{r5}) can be deduced.

Next, we use parameters $\rho$ and $\delta$ to represent the bound of users' valuations. By using these two parameters, we can calculate the competitive ratio more simply and beautifully.
Considering some other costs such as software and manage fees, SaaS providers are willing to provide services only if users' valuations are larger than the cost of purchasing instances.
Hence, for most of users, their minimum valuations are no smaller than respective costs, which means the sum of all users' minimum valuations is larger than or equal to $\rho$ times the total instance cost:
\begin{equation}\label{assume1}
\sum_{i\in\mathcal{U}}v_{i}^{\mathit{min}}\geq \rho\sum_{i\in\mathcal{U}}\hat{\phi_{i}^{e^{*}}}
\end{equation}
Meanwhile, the maximum valuations that users are willing to pay are finite, which means the sum of all users' maximum valuations is smaller than or equal to $\delta$ times the total instance cost:
\begin{equation}\label{assume2}
\sum_{i\in\mathcal{U}}v_{i}^{\mathit{max}}\leq \delta\sum_{i\in\mathcal{U}}\hat{\phi_{i}^{e^{*}}}
\end{equation}
$\rho$ and $\delta$ can be obtained by historical data.

Finally, we calculate the competitive ratio of the revenue $\mathit{REV}$ as follows:
\begin{align}\label{ratio_REV}
E[\frac{\mathit{REV_{opt}}}{\mathit{REV_{online}}}]
&\leq E[\frac{\sum_{i\in \mathcal{U}}\!v_{i}^{\mathit{max}}}{\sum_{i\in \mathcal{U}}\!v_{i}^{\mathit{min}} \!- \!2\!\sum_{i\in \mathcal{U}}\!\hat{\phi_{i}^{e^{*}}}}] \\
&\leq E[\frac{\delta\sum_{i\in \mathcal{U}}\hat{\phi_{i}^{e^{*}}}}{\rho\sum_{i\in \mathcal{U}}\hat{\phi_{i}^{e^{*}}} - 2\sum_{i\in \mathcal{U}}\hat{\phi_{i}^{e^{*}}}}]\label{r6}\\
&=\frac{\delta}{\rho-2} \label{ratio_REV}
\end{align}
Considering equations (\ref{assume1}) and (\ref{assume2}), equation (\ref{r6}) can be deduced.

In conclusion, Theorem \ref{thm3} can be proved.

\end{proof}

\begin{thm} \label{thm2}
The proposed online algorithm can achieve an expected competitive ratio of $\frac{\delta}{\theta(\rho-2)}$ in the aggregated objective $\mathit{OBJ}$. $\delta$ and $\rho$ are related to users' valuations. $\theta \in [0, 1]$. 
\end{thm}

\begin{proof}
Considering the same fictional situation mentioned above, according to equations (\ref{rev_opt}) and (\ref{sat_opt}), we can calculate the optimal solution $OBJ_\mathit{{opt}}$ as follows:
\begin{align}
OBJ_\mathit{{opt}} &=\theta\cdot\frac{\mathit{REV_{opt}}}{\mathit{REV^{*}}}+(1-\theta)\cdot\mathit{SAT_{opt}} \nonumber \\
&\leq \theta \cdot\frac{\sum_{i\in \mathcal{U}}v_{i}^{\mathit{max}}}{\mathit{REV^{*}}}+(1-\theta)\cdot 1 \nonumber\\
&\leq \frac{\sum_{i\in \mathcal{U}}v_{i}^{\mathit{max}}}{\mathit{REV^{*}}} \label{a1}
\end{align}
Since $\mathit{REV_{opt}}\leq\mathit{REV^{*}}$, $\frac{\mathit{REV_{opt}}}{\mathit{REV^{*}}}\geq 1$.
Thus equation (\ref{a1}) can be deduced.

For the $\mathit{OBJ_{online}}$, $E[\mathit{REV_{online}}]=E[\sum_{i\in \mathcal{W}}v_{i}^{e^{*}} - f(\mathcal{W})]$ (equation (\ref{r1})) and $\mathit{SAT_{online}}=\frac{n_{1}}{n}$.
Then the expectative aggregated objective $\mathit{OBJ_{online}}$ of our algorithm is:
\begin{align*}
E[&\mathit{OBJ_{online}}]=E[\theta\cdot\frac{\sum_{i\in \mathcal{W}}v_{i}^{e^{*}} - f(\mathcal{W})}{\mathit{REV^{*}}}+(1-\theta)\cdot\frac{n_{1}}{n}]\\
&\geq E[\theta\cdot\frac{\sum_{i\in \mathcal{U}}v_{i}^{e^{*}} - \sum_{i\in \mathcal{L}}v_{i}^{e^{*}} - \sum_{i\in\mathcal{U}}\hat{\phi_{i}^{e^{*}}}}{\mathit{REV^{*}}}+(1-\theta)\!\cdot\!\frac{n_{1}}{n}]\\
&\geq E[\theta\cdot\frac{\sum_{i\in \mathcal{U}}v_{i}^{e^{*}} - \sum_{i\in \mathcal{L}} (\hat{\phi_{i}^{e^{*}}}-\mathit{REV^{*}}\frac{1-\theta}{\theta}\frac{1}{n})-\sum_{i\in\mathcal{U}}\hat{\phi_{i}^{e^{*}}}} {\mathit{REV^{*}}}+(1-\theta)\cdot\frac{n_{1}}{n}]\\
&\geq E[\theta\cdot\frac{\sum_{i\in \mathcal{U}}v_{i}^{e^{*}} - 2\sum_{i\in\mathcal{U}}\hat{\phi_{i}^{e^{*}}}}{\mathit{REV^{*}}}+(1-\theta)]\\
&\geq E[\theta\cdot\frac{\sum_{i\in \mathcal{U}}v_{i}^{\mathit{min}}- 2\sum_{i\in \mathcal{U}}\hat{\phi_{i}^{e^{*}}}}{\mathit{REV^{*}}}]
\end{align*}
The main process is similar to the proof of $\mathit{REV_{online}}$, i.e., equations (\ref{r1})-(\ref{r5}).

Then the competitive ratio of the aggregated objective $\mathit{OBJ}$ is:
\begin{align}\label{ratio_OBJ}
E[\frac{\mathit{OBJ_{opt}}}{\mathit{OBJ_{online}}}]\nonumber
&\leq E[\frac{\sum_{i\in \mathcal{U}}v_{i}^{\mathit{max}}}{\mathit{REV^{*}}}\cdot\frac{\mathit{REV^{*}}}{\theta\cdot(\sum_{i\in \mathcal{U}}v_{i}^{\mathit{min}} - 2\sum_{i\in \mathcal{U}}\hat{\phi_{i}^{e^{*}}})}] \nonumber\\
&=\frac{\sum_{i\in \mathcal{U}}v_{i}^{\mathit{max}}}{\theta\cdot(\sum_{i\in \mathcal{U}}v_{i}^{\mathit{min}} - 2\sum_{i\in \mathcal{U}}\hat{\phi_{i}^{e^{*}}})}\nonumber\\
&\leq E[\frac{\delta\sum_{i\in \mathcal{U}}\hat{\phi_{i}^{e^{*}}}}{\theta(\rho\sum_{i\in \mathcal{U}}\hat{\phi_{i}^{e^{*}}} - 2\sum_{i\in \mathcal{U}}\hat{\phi_{i}^{e^{*}}})}]\nonumber\\
&=\frac{\delta}{\theta(\rho-2)}
\end{align}

In conclusion, Theorem \ref{thm2} can be proved.

\end{proof}

\begin{thm} \label{thm4}
The proposed online algorithm has polynomial time complexity.
\end{thm}

\begin{proof}
For each possible execution time of each user, the time complexity of calculating the instance purchasing and job scheduling schemes is $O(n+mT)$.
Since there is at most $O(nE)$ possible execution time, the time complexity of Algorithm \ref{online} is $O(n(n+mT)E)$. The process of re-computing does not increase the time complexity.

\end{proof}

\section{Experimental Evaluation}
In this section, we study the effectiveness and efficiency of our online algorithm through synthetic and Google data \cite{Googledata}. For SaaS provider's revenue $\mathit{REV}$ (equation (\ref{REV})), user satisfaction rate $\mathit{SAT}$ (equation (\ref{SAT})), and aggregated objective $\mathit{OBJ}$ (equation (\ref{OBJ})), our algorithm achieves great performance under various scenarios.

\subsection{Simulation Setup}

We simulate a cloud environment where a SaaS provider rents instances to provide pleasingly parallel job-execution services to users.
In practice, users can submit their jobs to the SaaS provider through a web page.
Then the SaaS provider schedules these jobs using our proposed online algorithm or other comparison algorithms on an IaaS cloud platform.
In this paper, we program a simulator and focus on the simulation of the presented mathematical model.
In our simulator, a simulated IaaS platform configures total $m\!=\!4$ heterogeneous instance types and allows on-demand accesses. The concrete configurations and prices of instances are set according to the real instances provided by Amazon EC2.
The job workloads used in our simulations are synthesized according to the experimental requirements, based on the observation of actual pleasingly parallel jobs and Google cluster data.
We try to emulate a real cloud. However, because of the constraints of experimental conditions, we simplify some details such as the bandwidth constraints of the network and the performance interference of inter-VMs.
The number of time slots $T$ is set based on experimental scenarios.
Each time slot consists of 10 minutes and thus the usage duration of a unit on-demand instance is $\tau=6$ slots.
Users' value functions decrease with the increase of execution time.
By default, we set $\alpha=0.5$ and $\theta=0.5$.
We randomly choose demand $D$ and threshold $k$ from specific ranges. The demand $D_{i,j}$ of job $i$ for type-$j$ instance is related to this instance's performance. The higher the performance is, the smaller the demand is.
For the experiments using Google data, we can get the demands and DoP thresholds from Google datasets, which contain the information of jobs submitted to Google cluster.

Table \ref{experiment} shows the concrete experimental parameters.
To obtain the optimal solution as a comparison, we relax the integer constraints of IP (\ref{IP}) and solve the linear programming, which generates the upper bound of the optimal solution.
To reduce the effect of randomness, all the experiments are repeated 20 times to generate average results.

\begin{table*}
\renewcommand{\arraystretch}{1.5}
\caption{Summary of Parameters}
\label{experiment}
\centering
\begin{tabular}{c|c|c|c|c|c|c|c}
\hline
\multirow{3}{*}{}& \multicolumn{7}{c}{Experiments} \\
\cline{2-8}
& \multicolumn{3}{c|}{EXP1} & EXP2 & EXP3 & EXP4 &EXP5 \\
\hline
\hline
\multirow{2}{*}{$n$} & Nor & Uni & Cons & \multicolumn{4}{c}{Nor} \\
\cline{2-8}
& \multicolumn{3}{c|}{2, 5, 8, ..., 26} & \multicolumn{2}{c|}{10} & 2, 5, 8, ..., 26 & 10 \\
\hline
$T$ &  \multicolumn{6}{c|}{200} & 60, 80, ..., 300\\
\hline
$D$ & \multicolumn{6}{c|}{Uniform [1, 30]} & Google dataset \\
\hline
$k$ & \multicolumn{6}{c|}{Uniform [5, 30]} & Google dataset\\
\hline
$E$ & \multicolumn{3}{c|}{6} & 2, 6, ..., 22 & \multicolumn{3}{c}{6} \\
\hline
\end{tabular}
\end{table*}

\subsection{EXP1: Performance with Different Numbers of Users per Time Slot}
We first study the performance of our online algorithm when the number of users per time slot increases.
To prove that our algorithm can adapt to multiple scenarios, we assume that user arrival follows normal (``Nor''), uniform (``Uni'') and constant (``Cons'') distributions respectively.
The results are shown in two aspects: (1) comparison of optimal, actual and theoretical $\mathit{OBJ}$ and $\mathit{REV}$, and (2) the actual competitive ratio.

\begin{figure*}[t]
\centering
\subfloat[Aggregated objective ($\mathit{OBJ}$)]{
\raggedleft
\includegraphics[width=4.2in]{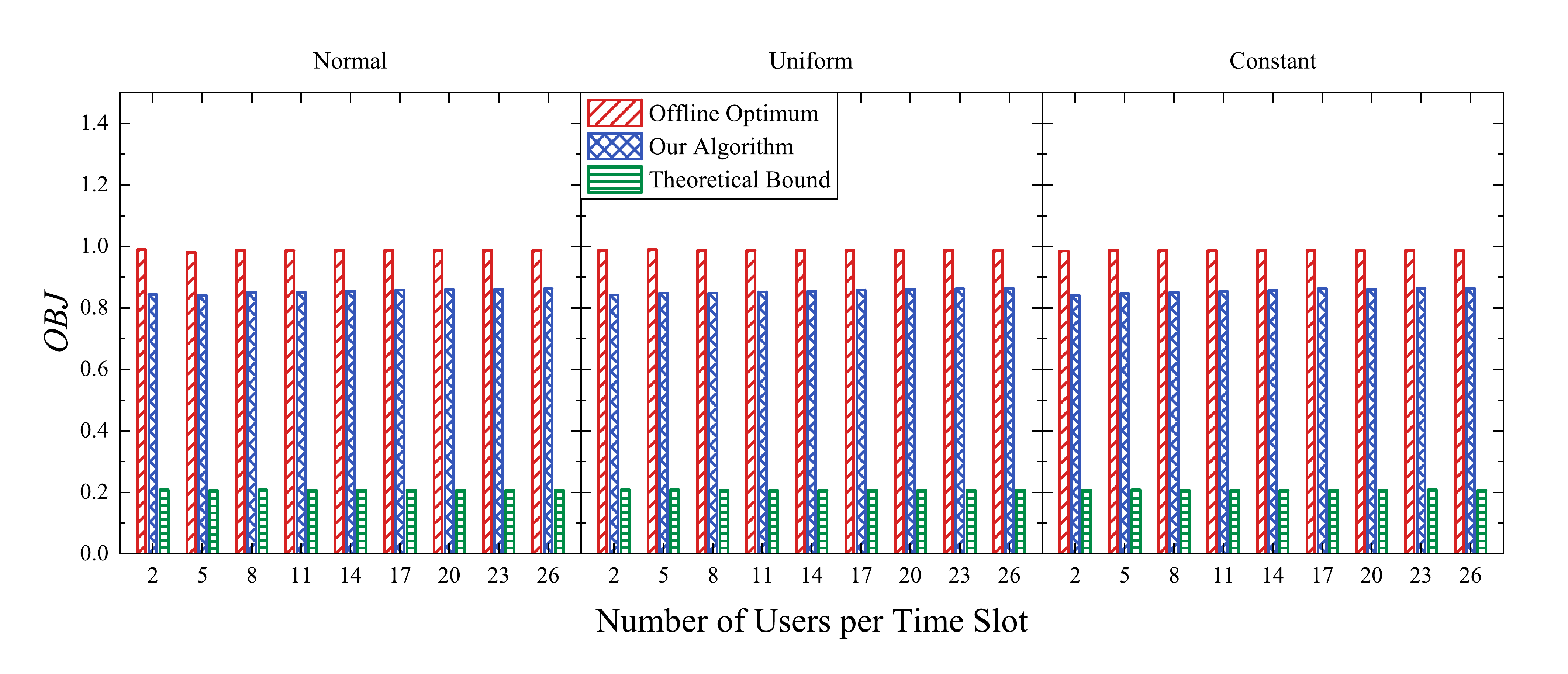}
\label{EXP1_OBJ}
}\
\subfloat[SaaS providers' revenue ($\mathit{REV}$)]{
\raggedright
\includegraphics[width=4.2in]{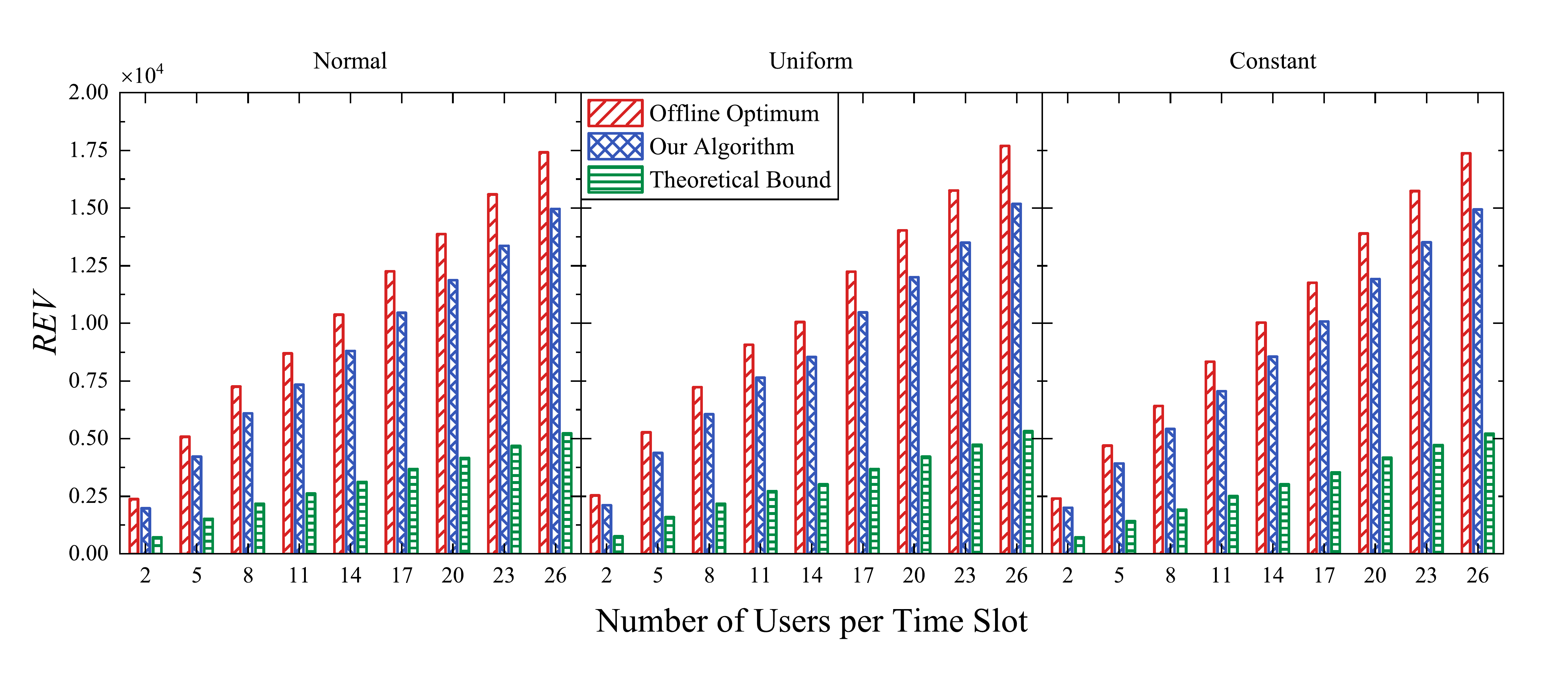}
\label{EXP1_REV}
}
\caption{Performance with different numbers of users per time slot.}
\label{EXP1}
\end{figure*}

\begin{figure*}[t]
\centering
\subfloat[Competitive ratio of $\mathit{OBJ}$]{
\raggedleft
\includegraphics[width=2.2in]{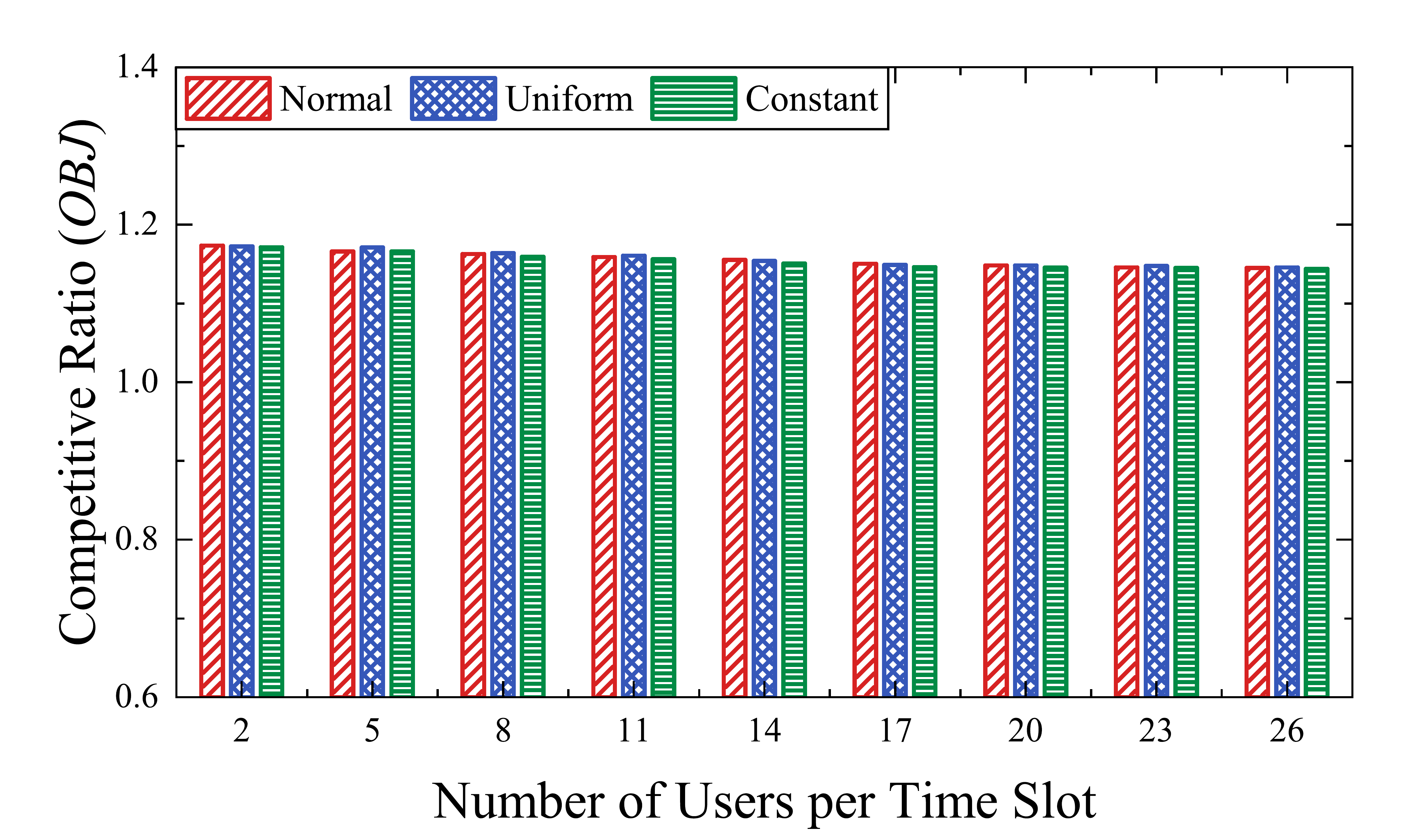}
\label{EXP2_OBJ}
}
\subfloat[Competitive ratio of $\mathit{REV}$]{
\raggedright
\includegraphics[width=2.2in]{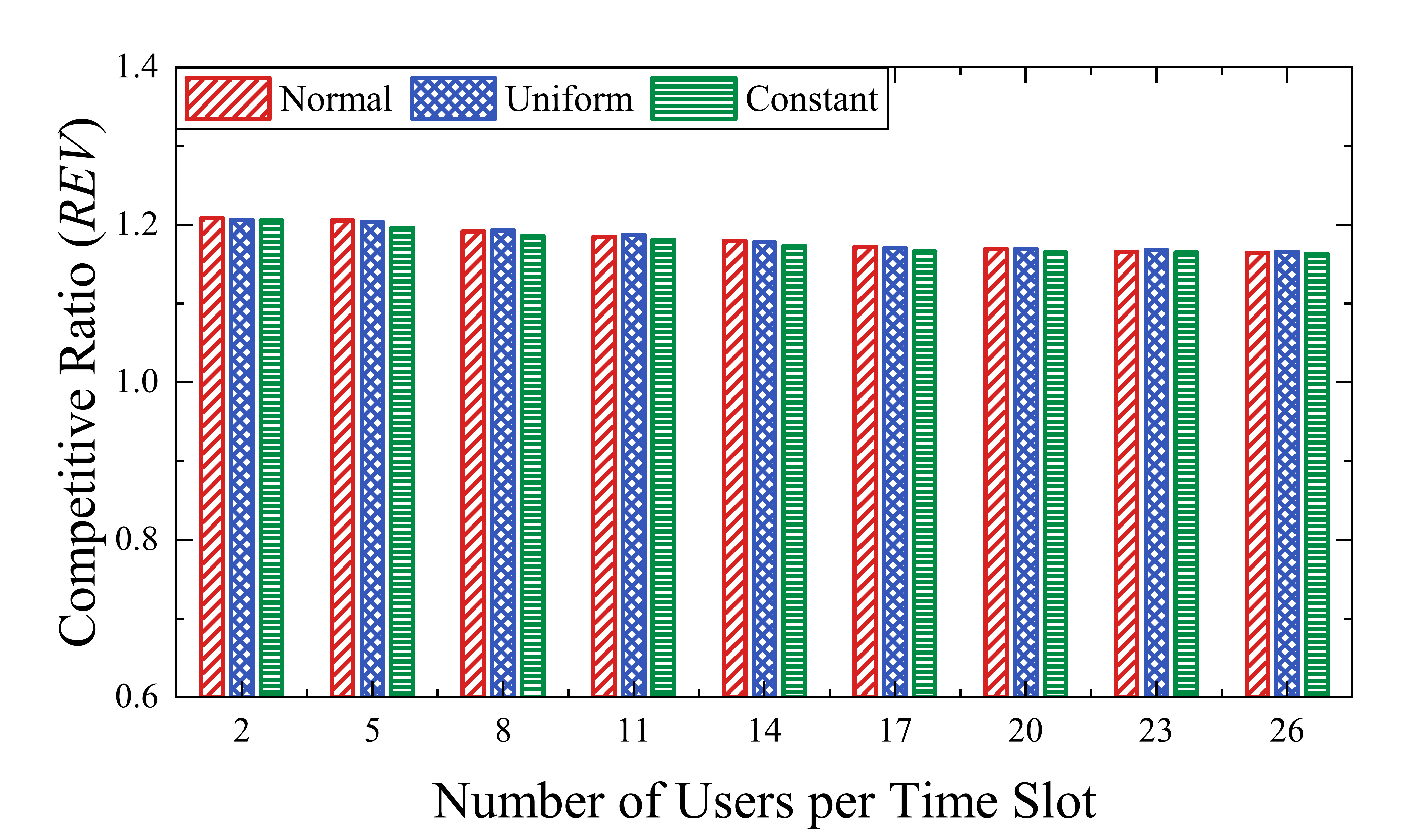}
\label{EXP2_REV}
}
\caption{Competitive ratio with different numbers of users per time slot.}
\label{EXP2}
\end{figure*}

From Fig. \ref{EXP1_OBJ}, we can see that as the number of users increases, the aggregated objective $\mathit{OBJ}$ is almost unchanged. This result reveals that our algorithm is stable under different numbers of users.
By contrast, in Fig. \ref{EXP1_REV}, SaaS providers' revenue $\mathit{REV}$ increases almost linearly, which means the more users, the more revenue. From this perspective, guaranteeing user satisfaction rate to attract users is important for SaaS providers.
Meanwhile, in both Fig. \ref{EXP1_OBJ} and Fig. \ref{EXP1_REV}, our algorithm's actual results are similar to optimal solutions, which also reflects the good performance.

According to equations (\ref{ratio_OBJ}) and (\ref{ratio_REV}), the theoretical competitive ratios of $\mathit{OBJ}$ and $\mathit{REV}$ in this experiment are 4.76 and 3.33 respectively.
It follows from Fig. \ref{EXP2_OBJ} and Fig. \ref{EXP2_REV} that the actual competitive ratios are very small. This is strong evidence that our algorithm can obtain great performance with different numbers of users. With the number of users per time slot increasing, the actual competitive ratios of both $\mathit{OBJ}$ and $\mathit{REV}$ show a slight decrease. This is due to the fact that with more users requesting services, a bad instance purchasing or scheduling scheme will produce less influence to the entire decision.
In addition, we can see that different distributions lead to similar results. This implies that our algorithm can be applied in a wide range of scenarios.

\subsection{EXP2: Performance with Different Numbers of Possible Execution Time}
We then study the performance of our online algorithm when the number of possible execution time increases.
The results are illustrated from the perspective of competitive ratio.

\begin{figure*}[t]
\centering
\subfloat[Competitive ratio of $\mathit{OBJ}$]{
\raggedleft
\includegraphics[width=2.2in]{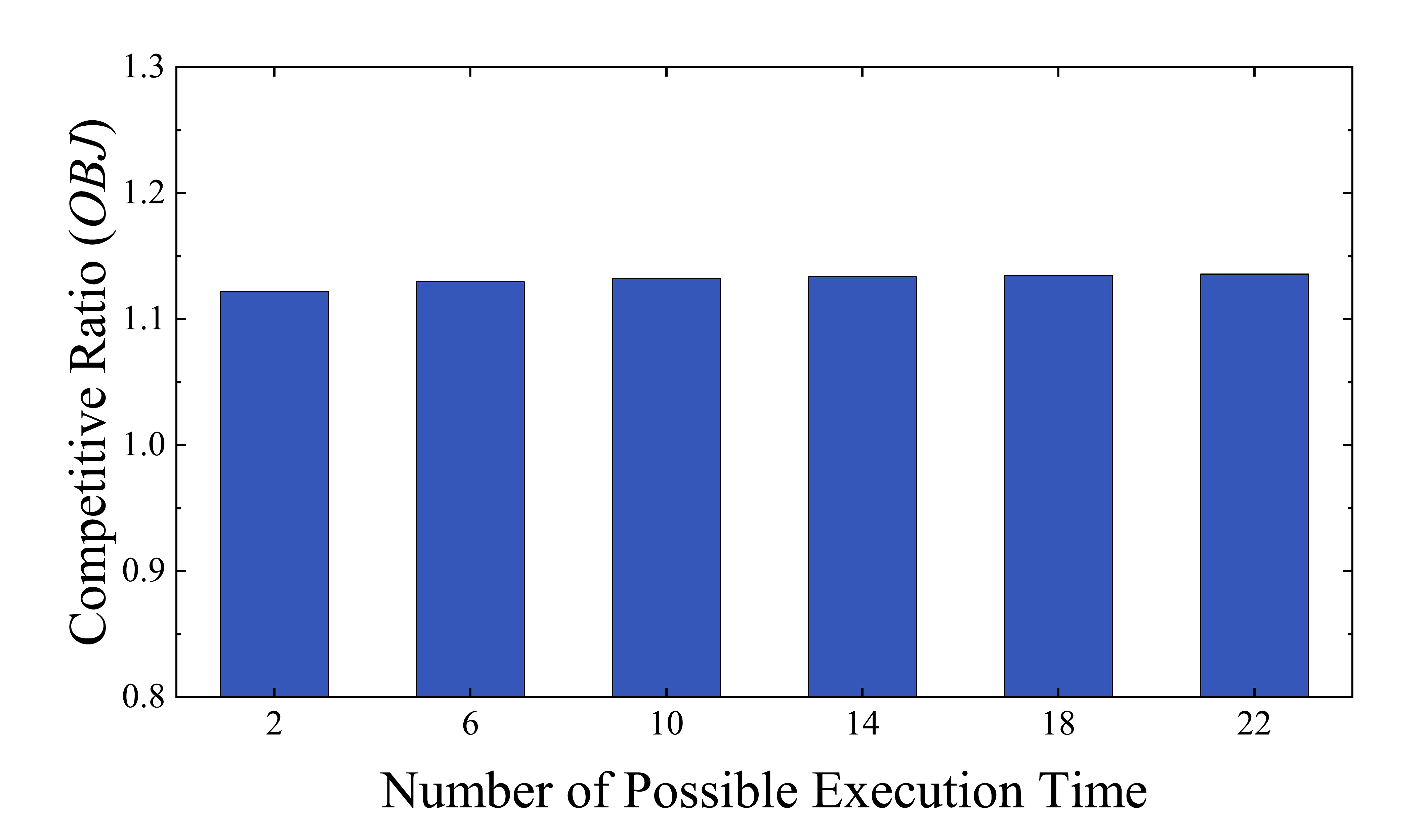}
\label{EXP3_OBJ}
}
\subfloat[Competitive ratio of $\mathit{REV}$]{
\raggedright
\includegraphics[width=2.2in]{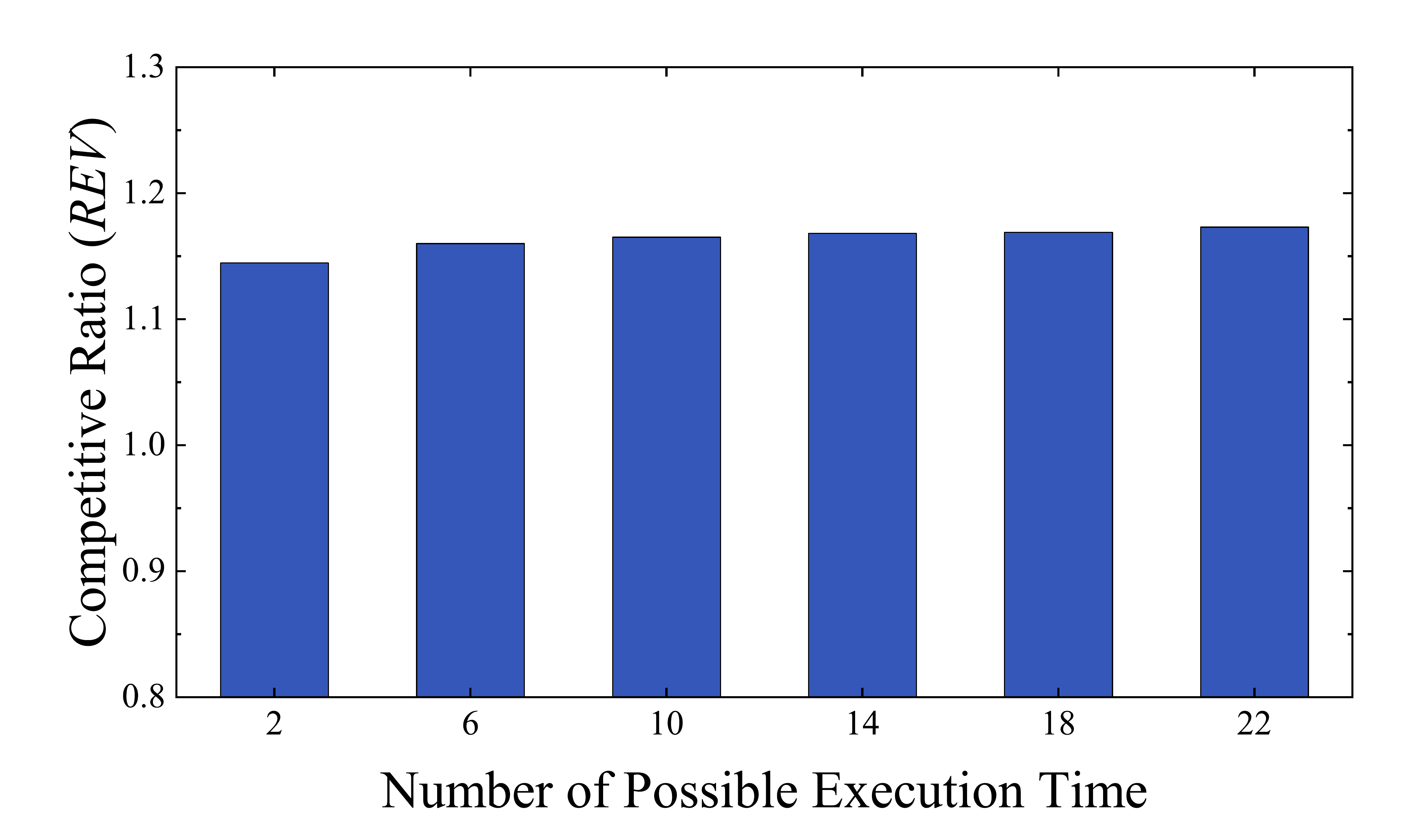}
\label{EXP3_REV}
}
\caption{Performance with different numbers of possible execution time.}
\label{EXP3}
\end{figure*}

From Fig. \ref{EXP3}, we can see that as the number of possible execution time increases, the competitive ratios of both $\mathit{OBJ}$ and $\mathit{REV}$ gradually increase. This is because as the number of possible execution time increases, out algorithm has more difficulties to make optimal decisions.
However, the competitive ratios are still very small, reflecting the good performance of our algorithm.

\subsection{EXP3: Performance with Different Values of $\theta$}
We next study the change of SaaS providers' revenue $\mathit{REV}$ and user satisfaction rate $\mathit{SAT}$ when the value of $\theta$ varies from 1 to 0.

\begin{figure*}[t]
\centering
\subfloat[SaaS providers' revenue ($\mathit{REV}$)]{
\raggedleft
\includegraphics[width=2.2in]{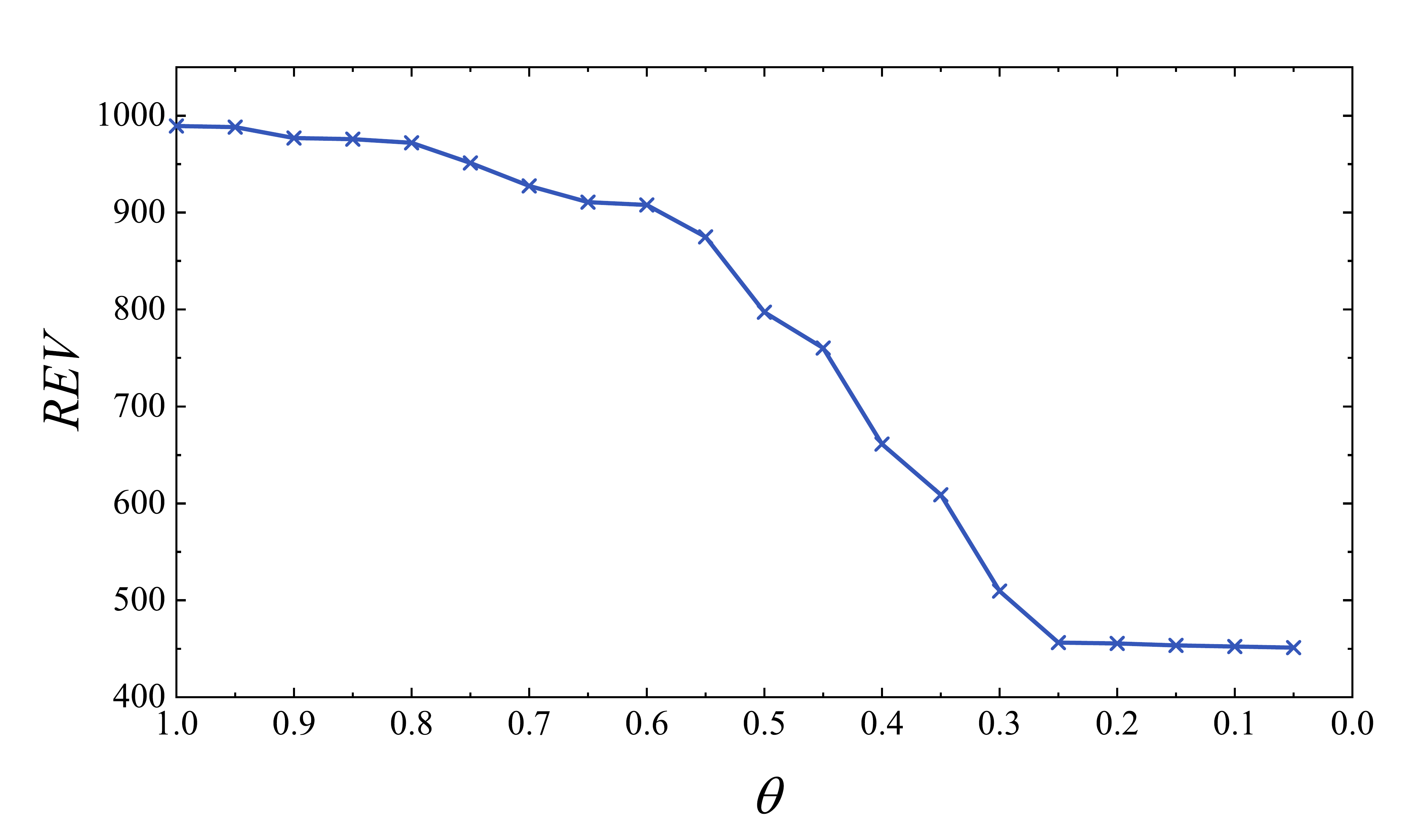}
\label{EXP4_OBJ}
}
\subfloat[User satisfaction rate ($\mathit{SAT}$)]{
\raggedright
\includegraphics[width=2.2in]{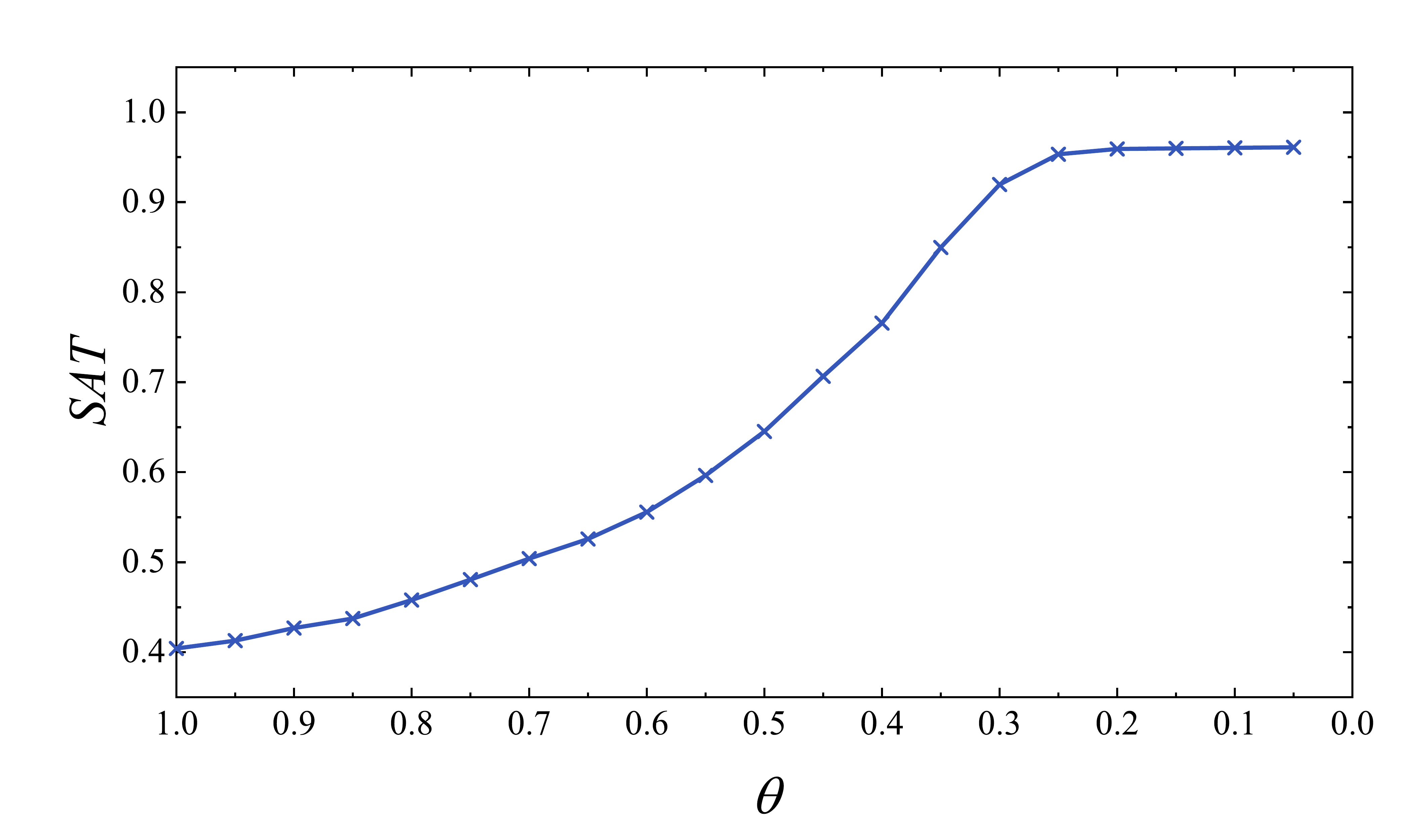}
\label{EXP4_REV}
}
\caption{Performance with different values of $\theta$.}
\label{EXP4}
\end{figure*}

It follows from Fig. \ref{EXP4} that with $\theta$ decreasing, providers' revenue gradually decreases and user satisfaction rate gradually increases.
This is because when $\theta$ is large, the algorithm will consider more about revenue. As $\theta$ decreases, the algorithm gradually focuses more on user satisfaction rate, leading to the shift of decisions.
When $\theta$ is small, the changes become slight.

\subsection{EXP4: Runtime}
To work efficiently in online markets, algorithms should run in polynomial-time.
We next compare the execution time of solving the optimal solution and our online algorithm, when the number of users per time slot increases.

\begin{figure}[!t]
\centering
\includegraphics[width=4.2in]{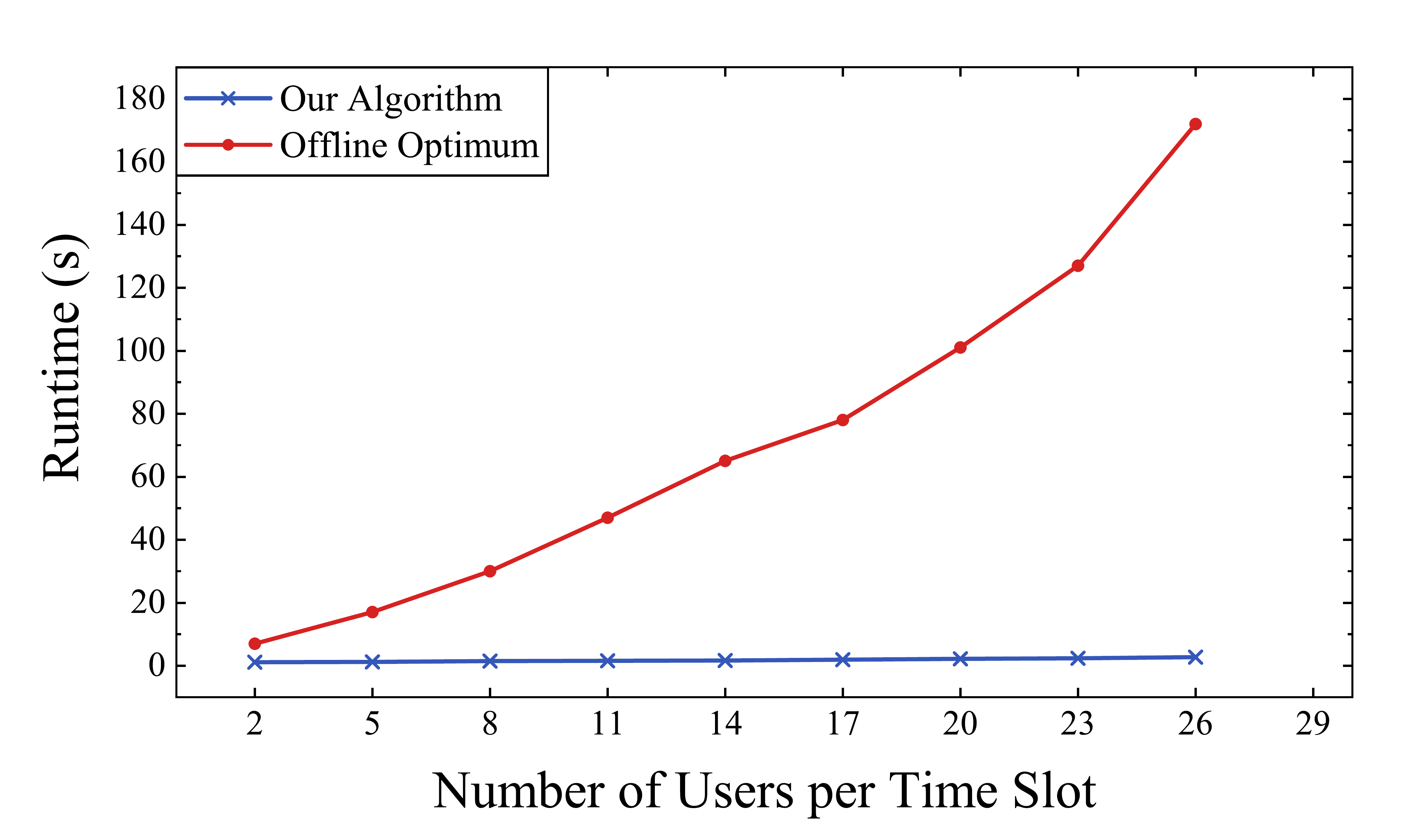}
\caption{Runtime.}
\label{EXP5}
\end{figure}

According to Fig. \ref{EXP5}, it is obvious that to achieve optimal results, the time cost is quite large. As the number of users increases, the runtime increases rapidly.
By contrast, our algorithm has polynomial time complexity and can be run much faster.
Combining former experimental results, one can conclude that our algorithm achieves good performance with a small time cost. Thereby, our algorithm can be used widely in practice.

\subsection{EXP5: Comparison with Other Algorithms}
Finally, we compare our online algorithm with some other online scheduling algorithms under Google data, when the number of time slots increases.

Currently, there are rare studies which consider the pleasingly parallel jobs, the variable resources, the online settings, and the complex objectives at the same time.
For comparison, we select several classic or recently published algorithms and make some adaptations to make them comparable with our proposed algorithm.
We compare our online algorithm with the following five scheduling algorithms by evaluating the aggregated objective $\mathit{OBJ}$, SaaS providers' revenue $\mathit{REV}$, and user satisfaction rate $\mathit{SAT}$:\\
(1) \emph{Primal-Dual}: The online primal-dual scheduling algorithm in \cite{Zhou2017An} focuses on pleasingly parallel jobs with the aim of revenue maximization. Their algorithm utilizes the dual programming of the problem and sets marginal price functions for resources. According to the valuations and prices, this algorithm decides whether to accept users' service requests.
Nevertheless, resource purchasing is not considered in this algorithm. Hence we assume that the number of purchased instances is smaller (``PD-small'') or larger (``PD-large'') than the total demand respectively.\\
(2) \emph{Earliest Deadline First}: This is one of the most classic scheduling algorithms \cite{Liu1973Scheduling, Dertouzos1974Control}. Jobs with the earliest deadline will be scheduled first. The resource purchasing in this algorithm is also not considered and we set fixed instances which are similar to users' demands. ``EDF'' is used to indicate this algorithm.\\
(3) \emph{Equal Opportunity}: This algorithm purchases instances, accepts users, and schedules jobs with equal opportunity. We use ``Equal-opp'' to indicate this algorithm for short.\\
(4) \emph{OnTaPRA}: This online algorithm designs for pleasingly parallel jobs and considers energy consumption \cite{2017Energy}. The main idea of this algorithm is: according to the \emph{Shortest Job First} policy, put the tasks on the instances with the highest efficiency as long as the constraints are not violated. In this way, the job completion time will be reduced. Meanwhile, the instance purchasing and cost are also not considered in this paper. Assuming that instances are fixed which are similar to users' demands and energy consumption is not considered, we modify this algorithm and apply it to our problem as a comparison.\\
(5) \emph{Dynalloc}: This algorithm considers to use self-owned, on-demand and spot instances at the same time to execute jobs, with the aim of cost optimization \cite{Wu2020Toward}. The main idea of this algorithm is distributing the workloads equally among the entire deadline and minimizing the integer instance hours of on-demand instances. In this process, the algorithm uses the self-owned resources first, then the spot instances, and last the on-demand instances. To compare with our online algorithm, we consider on-demand instances only and schedule the jobs by the Shortest Job First policy when using this algorithm.

\begin{figure*}[t]
\centering
\subfloat[Aggregated objective ($\mathit{OBJ}$)]{
\raggedleft
\includegraphics[width=4.2in]{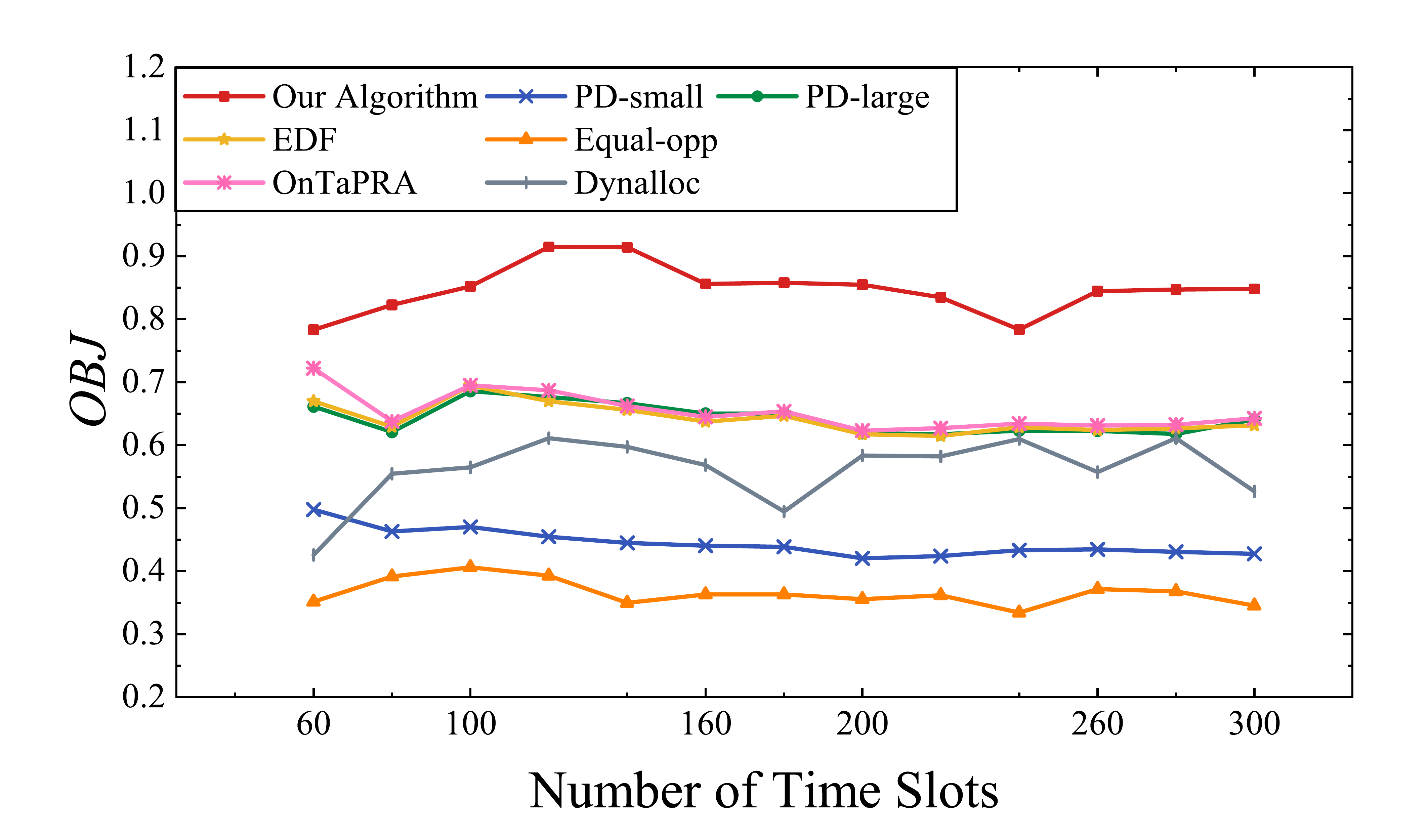}
\label{EXP6_OBJ}
}\
\subfloat[SaaS providers' revenue ($\mathit{REV}$)]{
\raggedleft
\includegraphics[width=2.2in]{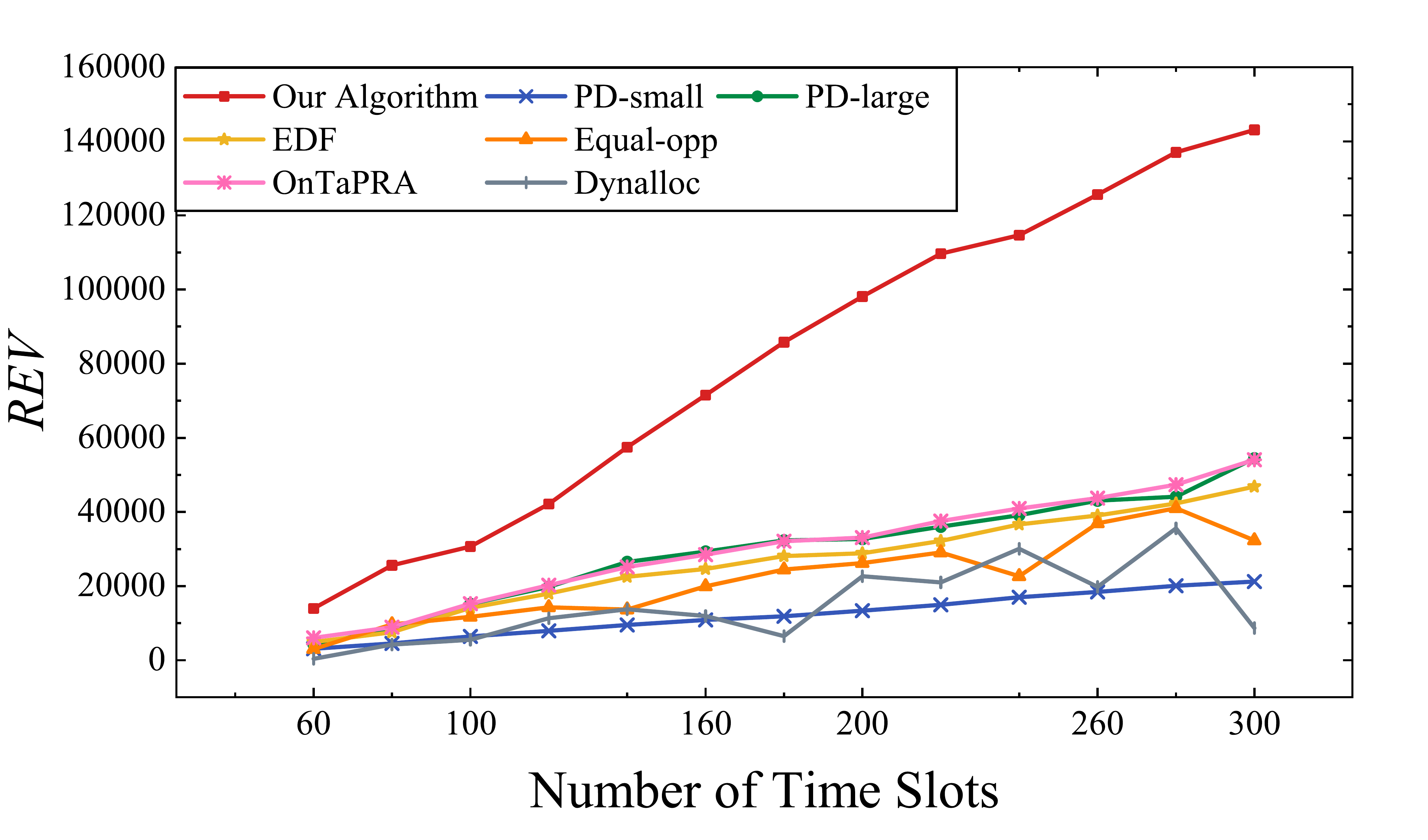}
\label{EXP6_REV}
}
\subfloat[User satisfaction rate ($\mathit{SAT}$)]{
\raggedright
\includegraphics[width=2.2in]{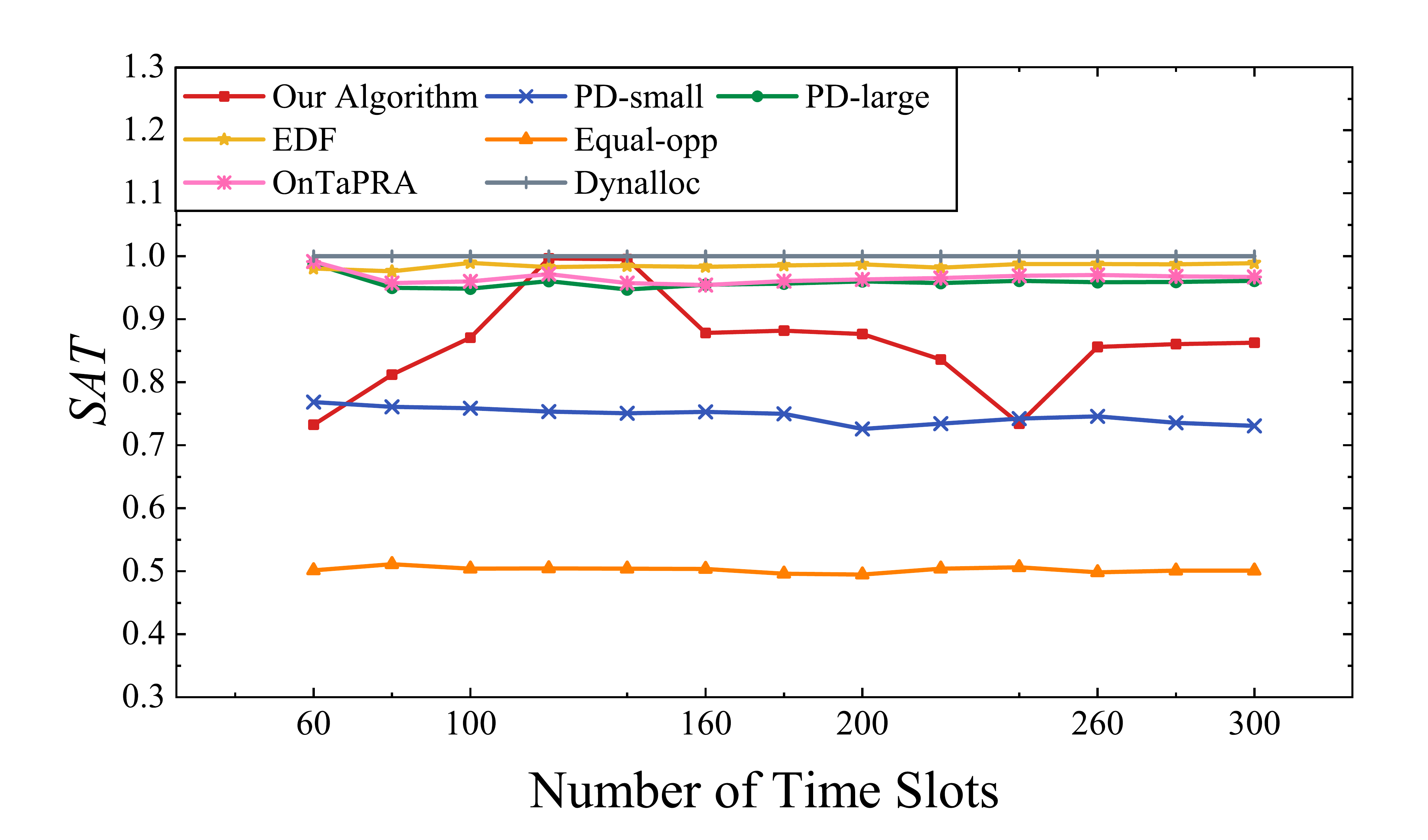}
\label{EXP6_SAT}
}
\caption{Comparison with other algorithms.}
\label{EXP6}
\end{figure*}

As shown in Fig. \ref{EXP6}, our online algorithm achieves better performance compared with all the comparison algorithms when applying for a long period.
In terms of $\mathit{OBJ}$ and $\mathit{REV}$, our online algorithm can achieve significantly better results compared to all the comparison algorithms. This is mainly caused by three reasons.
First, variable resources. Algorithms \emph{PD}, \emph{EDF} and \emph{OnTaPRA} assume fixed resources before the scheduling, which are actually hard to predict in advance. On the one hand, insufficient instances will limit the service capability of SaaS providers, such as \emph{PD-small}. On the other hand, excessive instances will produce insufferable costs, such as \emph{PD-large}. Thereby, SaaS providers should purchase instances dynamically according to real-time service requests rather than purchase a fixed number in advance.
Second, the scheduling objectives. Objectives decide the acceptance and scheduling policies of algorithms.
Algorithm \emph{PD} only considers the objective of revenue maximization and ignores the user satisfaction rate. The goal of algorithm \emph{OnTaPRA} is minimizing completion time.
Algorithms \emph{Equal-opp} and \emph{Dynalloc} can purchase resources dynamically. However, they ignore the economic factors (e.g., valuations of jobs and instance costs) and serve users as much as possible rather than accept and schedule jobs according to $\mathit{OBJ}$. Thus, there is a money loss.
Third, the online settings. In practical cloud environments, when making decisions, the future information about user arrivals and demands is unknown, which means current decisions could have good or bad implications for the future.
Thus when designing algorithms, we need to consider these uncertainties fully and carefully.
In the comparison algorithms, only \emph{PD} and \emph{OnTaPRA} consider the online settings.
All these comparison algorithms cannot fit well with our problem.

In terms of $\mathit{SAT}$, algorithms \emph{PD-large}, \emph{EDF}, \emph{OnTaPRA} and \emph{Dynalloc} obtain a larger user satisfaction rate. This is due to the fact that they have abundant instances to serve most users rather than purchasing instances and scheduling jobs selectively according to the objective $\mathit{OBJ}$. As a result, their $\mathit{OBJ}$ and $\mathit{REV}$ are lower than our online algorithm.

In summary, our online algorithm achieves better performance than all the five comparison algorithms and it can effectively deal with the market-oriented online bi-objective scheduling problem for pleasingly parallel jobs with variable resources in cloud environments.

\section{Conclusions and Future Work}
This paper focuses on the market-oriented online bi-objective scheduling problem for pleasingly parallel jobs with variable resources in cloud environments.
With unique features of variable resources and online settings etc., the scheduling focuses on bi-objective: maximizing SaaS providers' revenue and maximizing user satisfaction rate.
To address this problem, we propose an algorithm to help SaaS providers to decide how to purchase instances and schedule jobs in an online way.
Our proposed online algorithm is computationally efficient and can achieve a good competitive ratio.
The efficiency is verified through theoretical analysis and extensive simulations.
Nowadays, most IaaS cloud platforms also support some other instances such as reserved and spot instances, which are suitable for different situations. In the future, we plan to take other instances into consideration and design efficient algorithms to solve scheduling problems in more complicated cloud environments.

\section*{Acknowledgment}
This work was supported by the National KeyResearch and Development Program of China [grant number 2018YFB1702903].


\bibliography{ref}

\end{document}